\documentclass{article}

\usepackage{arxiv}

\usepackage{textpos}
\usepackage{enumitem}
\usepackage{rotating}
\usepackage[numbers,sort&compress]{natbib} 
\usepackage{tabularx,longtable,multirow,subfigure,caption,makecell}
\usepackage{fncylab} 
\usepackage{fancyhdr} 
\usepackage{url} 
\usepackage[english]{babel}
\usepackage{amsmath}
\usepackage{amsfonts}
\usepackage{graphicx}
\usepackage{siunitx}
\usepackage{dsfont}
\usepackage{nccmath}
\usepackage{titlesec}
\usepackage{lipsum}
\usepackage{epstopdf} 
\usepackage{array}
\usepackage{latexsym}
\usepackage{booktabs}
\usepackage[table,xcdraw]{xcolor}
\usepackage[normalem]{ulem}
\useunder{\uline}{\ul}{}
\usepackage{anyfontsize}
\usepackage[pdftex,hypertexnames=false,colorlinks]{hyperref} %
\usepackage{setspace} 
\hypersetup{pdftitle={},
  pdfsubject={}, 
  pdfauthor={},
  pdfkeywords={}, 
  pdfstartview=FitH,
  pdfpagemode={UseOutlines},
  bookmarksnumbered=true, bookmarksopen=true, colorlinks,
    citecolor=black,%
    filecolor=black,%
    linkcolor=black,%
    urlcolor=black}
\usepackage[all]{hypcap}
\usepackage{navigator}

\title{Transformers versus LSTMs for electronic trading}

\author{
    \anchor{https://orcid.org/0000-0001-6846-6649}{\includegraphics[scale=0.06]{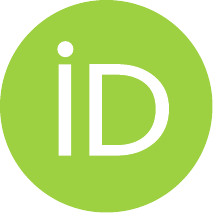}\hspace{1mm}Paul Bilokon} \\
	Department of Computing \\
	Imperial College London \\
	South Kensington Campus \\
	London SW7 2AZ \\
	\texttt{paul.bilokon@imperial.ac.uk}
	\And
	\anchor{https://orcid.org/0009-0005-1206-1710}{\includegraphics[scale=0.06]{orcid.pdf}\hspace{1mm}Yitao Qiu} \\
	Department of Computing \\
	Imperial College London \\
	South Kensington Campus \\
	London SW7 2AZ \\
	\texttt{yitao.qiu21@imperial.ac.uk}}

\begin{document}

\maketitle

\setlength{\abovedisplayskip}{3pt}
\setlength{\belowdisplayskip}{3pt}

\begin{abstract}
With the rapid development of artificial intelligence, long short term memory (LSTM), one kind of recurrent neural network (RNN), has been widely applied in time series prediction.\\\par

Like RNN, Transformer is designed to handle the sequential data. As Transformer achieved great success in Natural Language Processing (NLP), researchers got interested in Transformer’s performance on time series prediction, and plenty of Transformer-based solutions on long time series forecasting have come out recently. However, when it comes to financial time series prediction, LSTM is still a dominant architecture. Therefore, the question this study wants to answer is: whether the Transformer-based model can be applied in financial time series prediction and beat LSTM.\\\par

To answer this question, various LSTM-based and Transformer-based models are compared on multiple financial prediction tasks based on high-frequency limit order book data. A new LSTM-based model called DLSTM is built and new architecture for the Transformer-based model is designed to adapt for financial prediction. The experiment result reflects that the Transformer-based model only has the limited advantage in absolute price sequence prediction. The LSTM-based models show better and more robust performance on difference sequence prediction, such as price difference and price movement.
\end{abstract}

\section{Introduction}

Financial time series prediction is a significant task in investing and market-making activities. The Efficient Market Hypothesis proposed by Eugene \cite{EMH} states that all the information of the asset’s inner value has already precisely and completely reflected on the asset price, and it is impossible to beat the market by financial prediction. However, whether the market is efficient is questionable because technical analysis \cite{TA} believes the financial market is the physical movement of price (or features derived from prices). The price information can be interpreted by waves and patterns that can repeat themselves, where it is possible to make profitable buy or sell decisions in advance \cite{alma991000618293301591,boroden2008fibonacci}. During the prediction, challenging factors are noise and volatile features because price information is generally non-linear and non-stationary \cite{alma991000193490801591}. Lots of models are proposed to solve the financial time series problem. A typical linear model for regression is Auto-Regressive Integrated Moving average (ARIMA) \cite{7046047} and its variations, which requires domain expertise to handcraft features. With the development of machine learning, Artificial Neural Networks (ANN) raises great interest because of their capability to extract more abstract features from data and find a hidden non-linear relationship without assumptions or human expertise by adding more parameters. Long short-term memory (LSTM), which is a special recurrent neural network (RNN) architecture that has been proven successful in the application of sequential data, is widely applied to handwriting recognition  \cite{DBLP:journals/corr/abs-1902-10525} and speech recognition \cite{https://doi.org/10.48550/arxiv.1610.09975}. Like RNN, the Transformer \cite{attention} is also used to handle the sequential data. Compared to LSTM, the Transformer does not need to handle the sequence data in order, which instead confers the meaning of the sequence by the Self-attention mechanism. \\\par
Applying LSTM and Transformer for financial time series prediction is a popular trend nowadays. Depending on the historical financial data, researchers usually make predictions for the future numerical prices, price difference, return or future price movement (rise, stationary, fall). Although LSTM and Transformer are applied in different aspects for this problem. There are mainly two research directions: 
\begin{enumerate}[label={\arabic*)}]
  \item Make predictions based on high-frequency Limit Order Book (LOB) data and its derived features, such as Volume Order Imbalance (VOI) and Trade Flow Imbalance (TFI) \cite{LSTMLOB,DeepLOB,LOBs2s,https://doi.org/10.48550/arxiv.1810.09965,Kolm2021DeepOF}.
  \item  Make predictions based on OHLC (Open, High, Low, Close) data and its derived financial indices, such as Relative Strength Index (RSI) and Moving average convergence divergence (MACD) \cite{prdictLSTM,CAO2019127,10.1371/journal.pone.0180944,8126078,FISCHER2018654,DBLP:journals/corr/abs-1911-09512,AttenLSTM,Zhang_2019,9731073,9538640,predictBERT}.
\end{enumerate}
Since 2017, the Transformer has been increasingly used for Natural Language Processing (NLP) problems. It produces more impressive results than RNN, such as machine translation \cite{https://doi.org/10.48550/arxiv.1806.06957} and speech applications \cite{Karita_2019}, replacing RNN models such as LSTM in NLP tasks. Recently, a surge of  Transformer-based solutions for less explored long time series forecasting problem has appeared \cite{SURVEY}. However, as for the financial time series prediction, LSTM remains the dominant architecture.\\\par
Whether Transformer-based methods can be the right solution for financial time series forecasting is a problem worth investigating. Therefore, this paper is going to compare the performance of different Transformer-based and LSTM-based methods on financial time series prediction problems based on LOB data and attempt to adapt new models based on Transformer and LSTM. The contributions of this study are summed up as follows:
\begin{enumerate}[label={\arabic*.}]
  \item Systematically compare Transformer-based and LSTM-based methods in different financial prediction tasks based on high frequency LOB data collected from Binance Exchange. Tasks include (1) mid-price prediction, (2) mid-price difference prediction and (3) mid-price movement prediction. 
  \item For the first and second tasks, comparisons are all conducted on previous LSTM-based and Transformer-based methods. In the first task, the Transformer-based method has around $10\%-25\%$ prediction error less than the LSTM-based method, but the prediction result quality is insufficient for trading. In the second task, the LSTM-based method performs better than the Transformer-based method, where its highest out-of-sample $R^2$ reaches around $11.5\%$.
  \item The most significant contribution of this study is in the last task, mid-price movement prediction. A new LSTM-based model named DLSTM is developed for this task, which combines LSTM and the time series decomposition method. This model achieves $63.73\%$ to $73.31\%$ accuracy and shows strong profitability and robustness in simulated trading, outperforming previous LSTM and Transformer-based methods. In addition, the architecture of previous Transformer-based methods is also changed in order to adapt movement prediction task. 
\end{enumerate}
The later parts of this study are structured as follows: The background and related work are introduced in Section~\ref{section_2}. Section~\ref{section_3} describes the formulation of three financial prediction tasks. Section~\ref{section_4} explains the details of previous Transformer-based and LSTM-based methods used for comparison. Section~\ref{section_5} is the analysis of experiment results. Please note that the details of the newly developed DLSTM model and the architecture changes of Transformer models are explained in Section~\ref{task3_res} in Section~\ref{section_5}. The study is organized in this way to provide readers with a better understanding of the relationship among three financial prediction tasks.\\\par
The \textbf{Source Code} of this study is available at: \url{https://github.com/772435284/transformers_versus_lstms_for_electronic_trading}
\paragraph{Acknowledgements} We would like to express our gratitude to Zhipeng Wang for constructive comments and suggestions.
\section{Background and Related work}\label{section_2}
Time series prediction has been applied in different financial activities. For example, the trader or market maker predicts the future price or the price movement of the assets so that he/she can design trading/market making strategies based on the prediction results to make profit from it. This part examines papers related to time series prediction using LSTM and Transformer. Most of them are related to financial time series prediction.

\subsection{LSTM-based Time Series Prediction Solutions}

LSTM has been widely utilized in financial time series prediction depending on OHLC data and its derived financial indices. Many works \cite{prdictLSTM,CAO2019127,10.1371/journal.pone.0180944,8126078,FISCHER2018654} present predicting stock prices as successful by LSTM. Bidirectional LSTM (BiLSTM) is applied to increase the prediction performance \cite{DBLP:journals/corr/abs-1911-09512}. With the rise of NLP, Sequence-to-Sequence Model (S2S) \cite{S2S} is proposed and applied in machine translation and question answering. It is now also applied in financial time series prediction by combining LSTM structure and attention mechanisms contributing to higher performance \cite{AttenLSTM,Zhang_2019}.\\\par
LSTM has successfully forecasted high-frequency data depending on the large datasets extracted from the limit order book (LOB). In terms of making predictions upon order book data, Convolution Neural Network (CNN) and LSTM are both in favour of research and sometimes they are combined. Sirignano et al. \cite{LSTMLOB} trained a universal model using LSTM to predict the LOB price movement by data from all stocks, which outperforms linear and non-linear models trained on a specific asset. Zhang et al. \cite{DeepLOB} combine CNN and LSTM to form a new deep neural network architecture called DeepLOB to predict future stock price movements in LOB data outperforming architecture only containing LSTM. Zhang et al. \cite{LOBs2s} also combine the DeepLOB architecture with Seq2Seq and Attention model to forecast multi-horizon future price movements in one forward procedure, which reduces the training effort and achieves better performance in long horizon prediction than DeepLOB. According to Tsantekidis et al. \cite{https://doi.org/10.48550/arxiv.1810.09965}, the structure of CNN-LSTM architecture is able to outperform other models on LOB price movement prediction because of its more stable behaviour. Kolm et al. \cite{Kolm2021DeepOF} use the OFI feature derived from the LOB to predict the future min-price return, where the CNN-LSTM structure still achieves the best performance.

\subsection{Transformer-based Time Series Prediction Solutions}

As Transformer makes a great contribution to NLP \cite{https://doi.org/10.48550/arxiv.2005.14165}, many advanced models are proposed, such as BERT and GPT-3, which now have more influence in time series prediction. As the canonical Self-attention mechanism has $O(L^2)$ time and memory complexity, many modifications have been made to the Transformer in order to adapt to the time series prediction problem to process the long sequence efficiently. There are many alternative Transformer models for long time series forecasting problem have been developed recently \cite{SURVEY}: LogTrans\cite{LogTrans}, Reformer \cite{Reformer}, Informer \cite{Informer}, Autoformer \cite{Autoformer}, Pyraformer \cite{liu2022pyraformer} and the recent FEDformer \cite{FEDFormer}.\\

Theses Transformer models mentioned above are mainly evaluated on non-financial datasets, such as electricity consumption, traffic usage, and solar energy dataset. They achieve a considerable performance increase in accuracy over the LSTM model.  Some works start applying Transformers for financial time series prediction. Hu \cite{9731073} uses a Temporal Fusion Transformer with support vector regression (SVR) and LSTM to predict stock price. Sridhar et al. \cite{9538640} predict the Dogecoin price through Transformer, which has superior performance compared to LSTM. Sonkiya et al. \cite{predictBERT} do sentiment analysis by BERT to generate the sentiment score, which is then combined with other financial indices and passed into the Generative Adversarial Network (GAN) for stock price prediction. These works \cite{9731073,9538640,predictBERT} use Transformer-based methods to make predictions based on OHLC data and the research on applying Transformer for LOB prediction is limited. \\\par
Overall, LSTM is applied broadly in financial time series prediction and has been tested on various datasets, while the Transformer is limited. Therefore, this study wants to apply these Transformer-based models to a wider region in financial time series prediction to compare their performance to LSTM.

\section{Financial Time Series Prediction Tasks Formulation} \label{section_3}
This study compares LSTM-based and Transformer-based methods among three financial prediction tasks based on LOB data. In this section, the basic concept of LOB will be first introduced, and then the formulation of three financial prediction tasks will be explained in detail. Three tasks are listed below:
\begin{itemize}
    \item Task 1: LOB Mid-Price Prediction
    \item Task 2: LOB Mid-Price Difference Prediction
    \item Task 3: LOB Mid-Price Movement Prediction
\end{itemize}

\subsection{Limit Order Book}

\begin{figure}[h]
    \centering
    \includegraphics[scale=1]{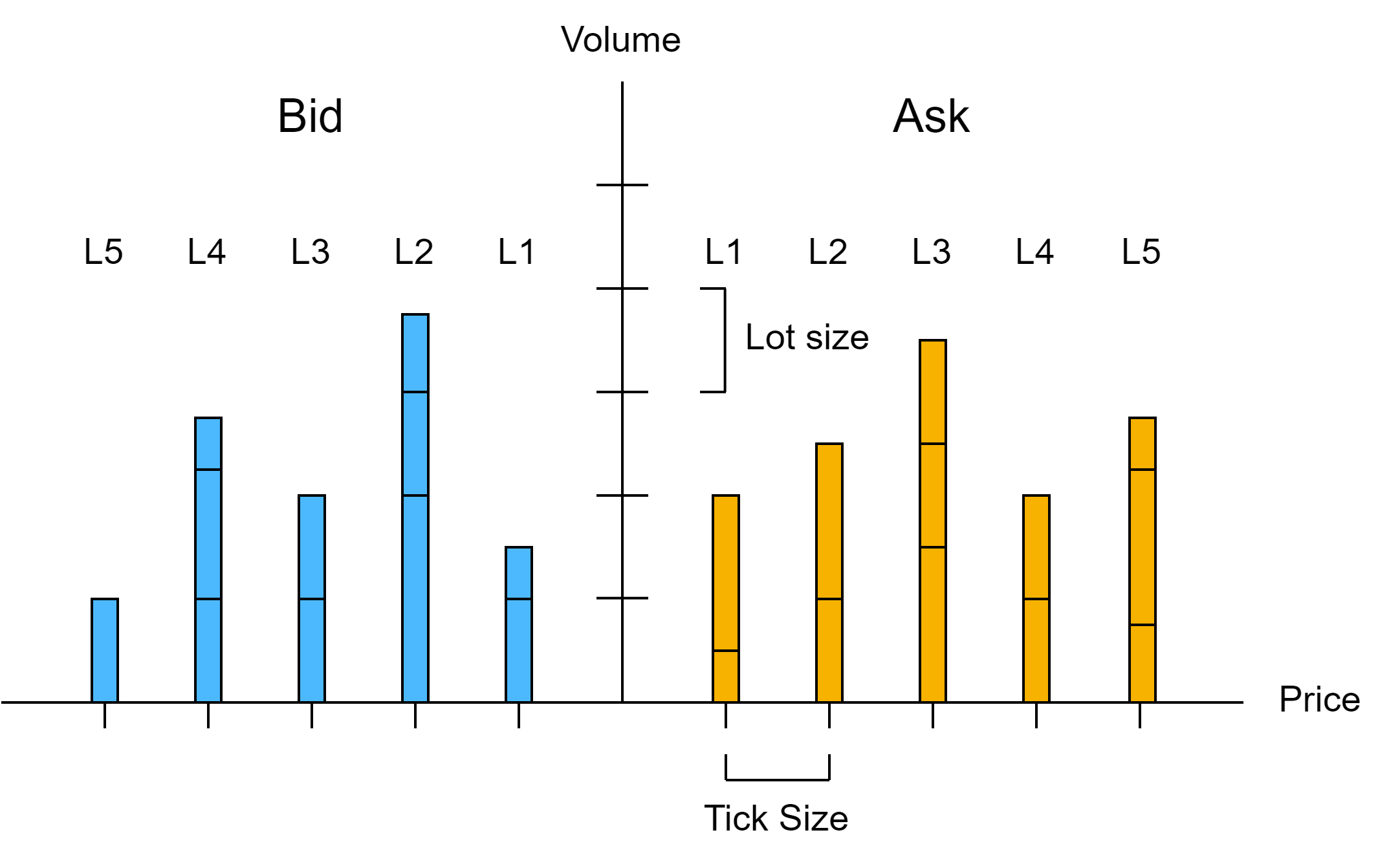}
    \caption{An illustration of LOB based on Zhang et al. \cite{DeepLOB} and Kolm et al. \cite{Kolm2021DeepOF}.}
    \label{fig.LOB}
\end{figure}

With computers and the Internet, most financial products such as stocks, forex and cryptocurrency are traded on the electronic market nowadays. Two types of orders exist in the electronic market: limit order and market order. According to Gould et al. \cite{LOB}, a limit order is the order to execute buy or sell direction at a specific price, where the orders can be succeeded, overdue, or cancelled and are recorded by the LOB. There are bid limit orders and ask limit orders used to buy and sell products by the trader or sell and buy products by the market marker. The highest bid price the buyers are ready to buy is referred to as the best bid price, and the lowest ask price the sellers are ready to sell is called the best ask price. The average of these two prices is called the mid-price, which reflects the current value of the financial product. The difference between them is the spread, which the market marker can usually make a profit. The illustration of the LOB is shown in Fig \ref{fig.LOB}.   
Another type of order is the market order, where the trader can immediately buy/sell the product at the best price. 
There exists a matching mechanism in the LOB. Most exchanges adopt the price and time priority matching mechanism. The limit orders will be first executed in order with a better price. If two orders have the same execution price, then the order that comes first in time will be executed first, following the first in first out (FIFO) principle. 

\subsection{Task 1: LOB Mid-Price Prediction}

The first task is to predict the LOB Mid-Price Prediction, which is to compare the ability to predict absolute price values similar to non-financial datasets in previous works \cite{LogTrans, Informer, Autoformer, FEDFormer,liu2022pyraformer}. The definition of time series prediction is given below and shown in Figure \ref{fig.tsf}:
\begin{figure}[h]
    \centering
    \includegraphics[scale=0.7]{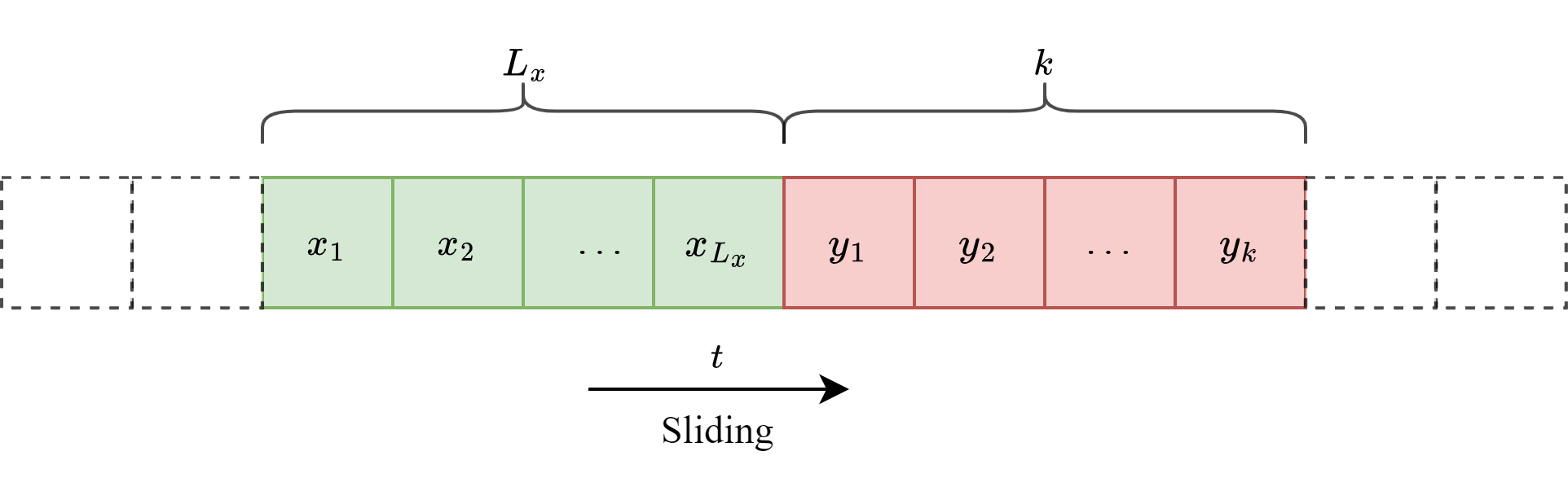}
    \caption{The illustration of time series prediction.}
    \label{fig.tsf}
\end{figure}\\
First, define a sliding window size $L_{x}$ for the past data. The input data at each time step $t$ is defined as:
\begin{equation}
    X_{t}=\left\{x_{1}, x_{2}, \ldots, x_{L_{x}}\right\}_{t}
\end{equation}
Then define a prediction window size $k$, where the goal is to predict the information in future $L_x+k$ steps. It will be the single-step prediction when $k=1$ and be multi-horizon prediction when $k>1$. Then the output at time step t is defined as:
\begin{equation}
    Y_{t}=\left\{y_{1}, y_{2}, \ldots, y_{k}\right\}_{t}
\end{equation}
The next step is to define the $x_t$ and $y_t$ in the input and output for mid-price prediction. Assume the market depth is 10. For a limit bid order at time t, the bid price is denoted as $p_{i, t}^{b i d}$ and the volume is $v_{i, t}^{b i d}$, where $i$ is the market depth. Same for the limit ask order, ask price is  $p_{i, t}^{ask}$ and volume is $v_{i, t}^{ask}$. Then the LOB data at time t is defined as:
\begin{equation}\label{eq:1}
    x_{t}=\left[p_{i, t}^{a s k}, v_{i, t}^{a s k}, p_{i, t}^{b i d}, v_{i, t}^{b i d}\right]_{i=1}^{n=10} \in R^{40}
\end{equation}
The past mid-price will be added to LOB data as input, and the mid-price is represented as:
\begin{equation}
    p_{t}^{\text {mid }}=\frac{p_{1, t}^{a s k}+p_{1, t}^{b i d}}{2}
\end{equation}
Finally, the $x_t$ will be:
\begin{equation}
    x_{t}=\left[p_{i, t}^{a s k}, v_{i, t}^{a s k}, p_{i, t}^{b i d}, v_{i, t}^{b i d},p_{t}^{\text {mid }}\right]_{i=1}^{n=10} \in R^{41}
\end{equation}
The target is to predict the future mid-price, so $y_t = p_{t}^{\text {mid }}$.

\subsection{Task 2: LOB Mid-Price Difference Prediction}\label{task2}

The second task is to predict the mid-price change, which is the the difference of two mid-prices in different time step. Trading strategies can be designed if the price change becomes negative or positive. The input of this task is the same as the mid-price prediction, as described in Equation \ref{eq:1}. The target is to regress the future difference between current mid-price $p_{t}^{\text {mid }}$ and the future mid-price $p_{t+\tau}^{\text {mid}}$:
\begin{equation}
    d_{t+\tau} = p_{t+\tau}^{\text {mid}} - p_{t}^{\text {mid }}
\end{equation}
Like the mid-price prediction, a prediction window size is defined as $k$, then the output of this task in each timestamp $t$  is represented as:
\begin{equation}
    Y_{t}=\left\{d_{t+1}, d_{t+2}, \ldots, d_{t+k}\right\}_{t}
\end{equation}

\subsection{Task 3: LOB Mid-Price Movement Prediction} \label{task3}

\begin{figure}[h]
    \centering
    \includegraphics[scale=0.42]{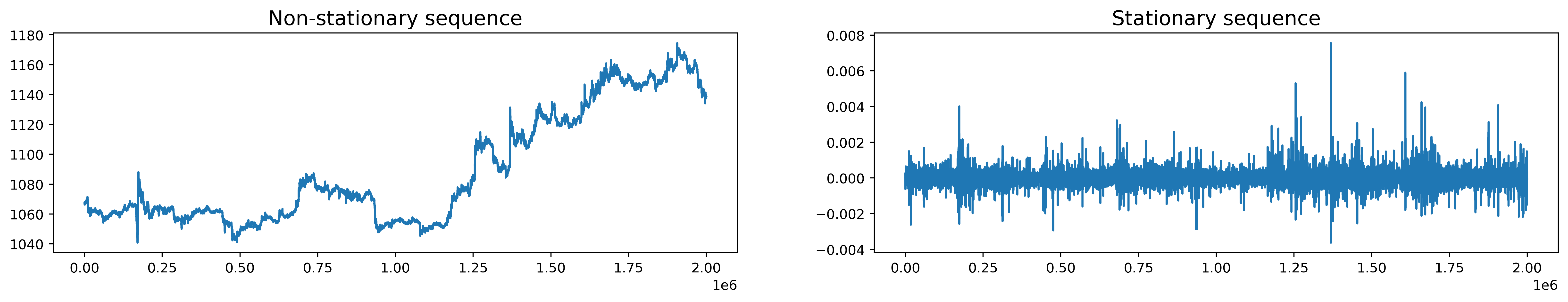}
    \caption{An example of non-stationary sequence vs stationary sequence.}
    \label{fig.s_ns}
\end{figure}

According to Ruppert \cite{alma991000193490801591}, the absolute price information is generally non-stationary, while the price change information, such as price difference and return and approximately stationary. The the mid-price difference is a difficult target for deep learning methods to predict because it is hard to extract meaningful pattern from it, although it helps design trading strategies. An example of non-stationary and stationary sequence is shown in Figure \ref{fig.s_ns}.

For this reason, an easier classification task for predicting mid-price movement is introduced here. To train a model to predict mid-price movement, the first step is to create price movement labels for each timestamp. This study follows the smoothing labelling method from Tsantekidis et al. \cite{8010701} and Zhang et al. \cite{DeepLOB}: Use $m^-$ to represent the average of the last $k$ mid-price and $m^+$ to represent the average of the next $k$ mid-price:
\begin{equation}
    m^{-}(t)=\frac{1}{k} \sum_{i=0}^{k} p_{t-k}^{m i d}
\end{equation}
\begin{equation}
    m^{+}(t)=\frac{1}{k} \sum_{i=1}^{k} p_{t+k}^{m i d}
\end{equation}
$k$ is set to $20, 30, 50, 100$ in this study following previous work of Zhang et al. \cite{DeepLOB}.\\
And then, define a percentage change $l_t$ to decide the price change direction.
\begin{equation}
    l_{t}=\frac{m^{+}(t)-m^{-}(t)}{m^{-}(t)}
\end{equation}
The label is dependent on the value of $l_t$. A threshold $\delta$ is set to decide the corresponding label. There are three labels for the price movement:
\begin{equation}
    \text { label }=\left\{\begin{array}{c}
0 (\text { fall }), \text { when } l_{t}>\delta \\
1 (\text { stationary }), \text { when }-\delta \leq l_{t} \leq \delta \\
2 (\text { rise }), \text { when } l_{t}<-\delta
\end{array}\right.
\end{equation}
\begin{figure}[hbt!]
    \centering
    \includegraphics[scale=0.45]{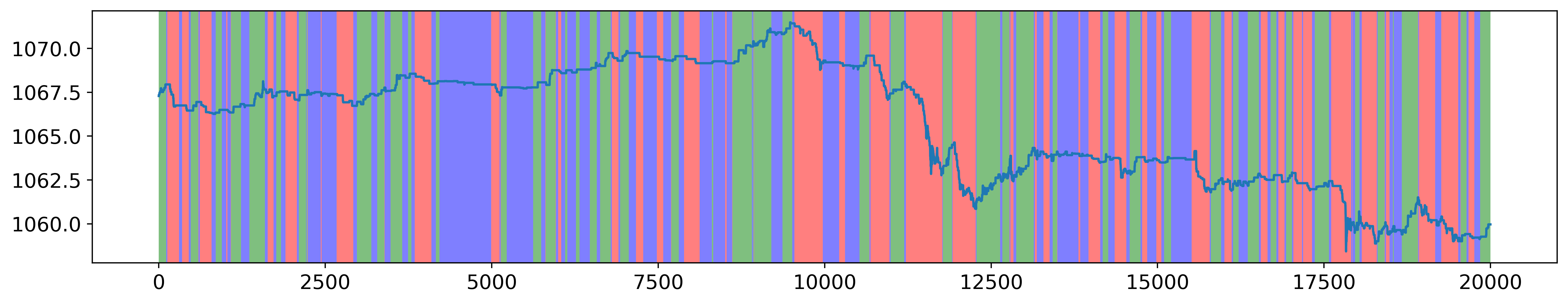}
    \caption{An example of labelling on horizon 100 on ETH-USDT dataset with fixed threshold $\delta$. The green colour represents the rise signal. Purple represents the price is stationary and red colour means the price fall.}
    \label{fig.labelling}
\end{figure}
An example of labelling for horizon 100 is shown in Figure \ref{fig.labelling}. 
Assume there is an input in Equation \ref{eq:1} at timestamp $t$, predicting mid-price movement is a one-step ahead prediction, which is to predict the mid-price movement in timestamp $t+1$.

\section{Methodology}\label{section_4}

\subsection{LSTM}

\begin{figure}[h]
    \centering
    \includegraphics[scale=0.7]{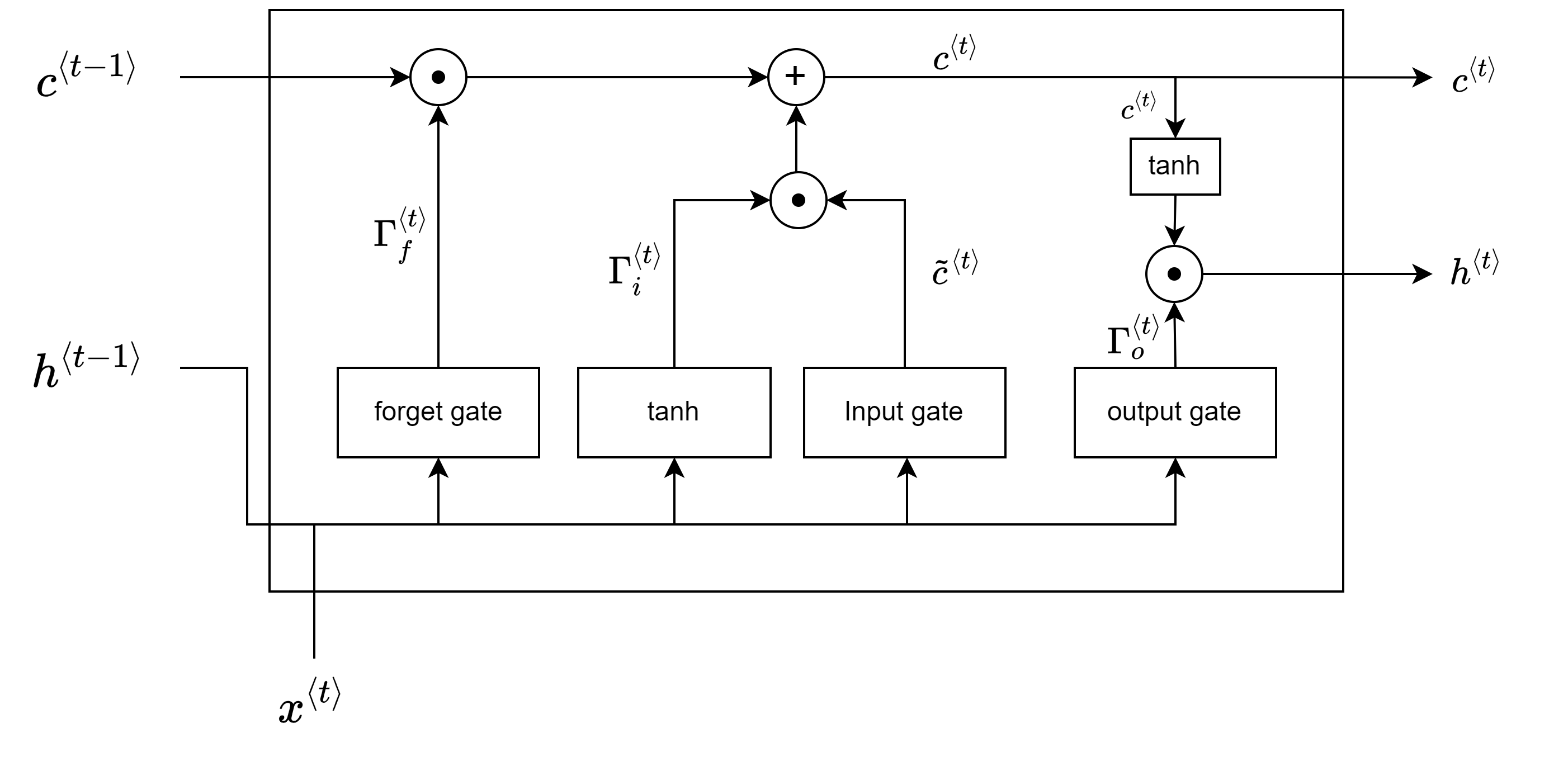}
    \caption{LSTM Structure based on Graves \cite{lstmfig1} and Fisher \cite{FISCHER2018654}.}
    \label{fig.lstm}
\end{figure}

LSTM was introduced by Hochreiter et al. \cite{lstm} is one of the RNNs with structural adaptability in time series data input. Although the traditional RNN has the capacity to store data, but it suffers from the exploding gradient problem and vanishing gradient problem. Exploding/vanishing gradient means the gradient that is used to update the neural networks increases/decreases exponentially, which makes the neural network untrainable \cite{dlbook}. Therefore, RNN is not successful in studying long-time series relations \cite{Rumelhart1986LearningRB}. LSTM neural network utilizes the coordination of three gates to keep long-term dependency and short-term memory. According to Gers et al. \cite{818041}, the three gates that LSTM utilizes are 1) forget gate, 2) input gate 3) output gate. The structure of an LSTM cell is shown in Figure \ref{fig.lstm}.
The calculations of the LSTM are as follows \cite{lstm}:\\
First, the LSTM need to decide to forget some information from the cell state, which is done by the forget gate. The forget gate has its sigmoid function $\sigma$. Then the function of forget gate is:
\begin{equation}
    \Gamma_{f}^{\langle t\rangle}=\sigma\left(W_{f}\left[h^{\langle t-1\rangle}, x^{\langle t\rangle}\right]+b_{f}\right)
\end{equation}
Where the $W_f$ is the weight of the last hidden state and input, $b_f$ is the bias of the hidden state.\\
The next step is to design what information the neural cell should remember. To update the information, the input gate should coordinate with a $tanh$ layer containing a vector of new candidate values $\tilde{c}^{\langle t\rangle}$. The calculation is as follows:
\begin{equation}
    \Gamma_{i}^{\langle t\rangle}=\sigma\left(W_{i}\left[h^{\langle t-1\rangle}, x^{\langle t\rangle}\right]+b_{i}\right)
\end{equation}
\useshortskip
\begin{equation}
    \tilde{c}^{\langle t\rangle}=\tanh \left(W_{c}\left[h^{\langle t-1\rangle}, x^{\langle t\rangle}\right]+b_{c}\right)
\end{equation}
And then, with the previous steps of calculations, the cell state can be updated:
\begin{equation}
    c^{\langle t\rangle}=\Gamma_{f}^{\langle t\rangle} \circ c^{\langle t-1\rangle}+\Gamma_{i}^{\langle t\rangle} \circ \tilde{c}^{\langle t\rangle}
\end{equation}
Lastly, calculate the result from the output gate to get the new hidden state:
\begin{equation}
    \Gamma_{o}^{\langle t\rangle}=\sigma\left(W_{o}\left[h^{\langle t-1\rangle}, x^{\langle t\rangle}\right]+b_{o}\right)
\end{equation}
\begin{equation}
    h^{\langle t-1\rangle}=\Gamma_{o}^{\langle t\rangle} \circ \tanh \left(\tilde{c}^{\langle t\rangle}\right)
\end{equation}
The output of LSTM can be every hidden state or the final hidden state, which depends on the application. In the implementation, this hidden state will be fed into a multi-layer perceptron (MLP), also known as feedforward-backwards propagation neural network (FFBPN). The output of this layer will pass through an activation function to generate the final output.
Usually, the final hidden state will be utilized in financial time series prediction, which produces absolute price prediction or price movement prediction.

\subsubsection{Alternative LSTM-based Models}

\begin{figure}[h]
    \centering
    \includegraphics[scale=1]{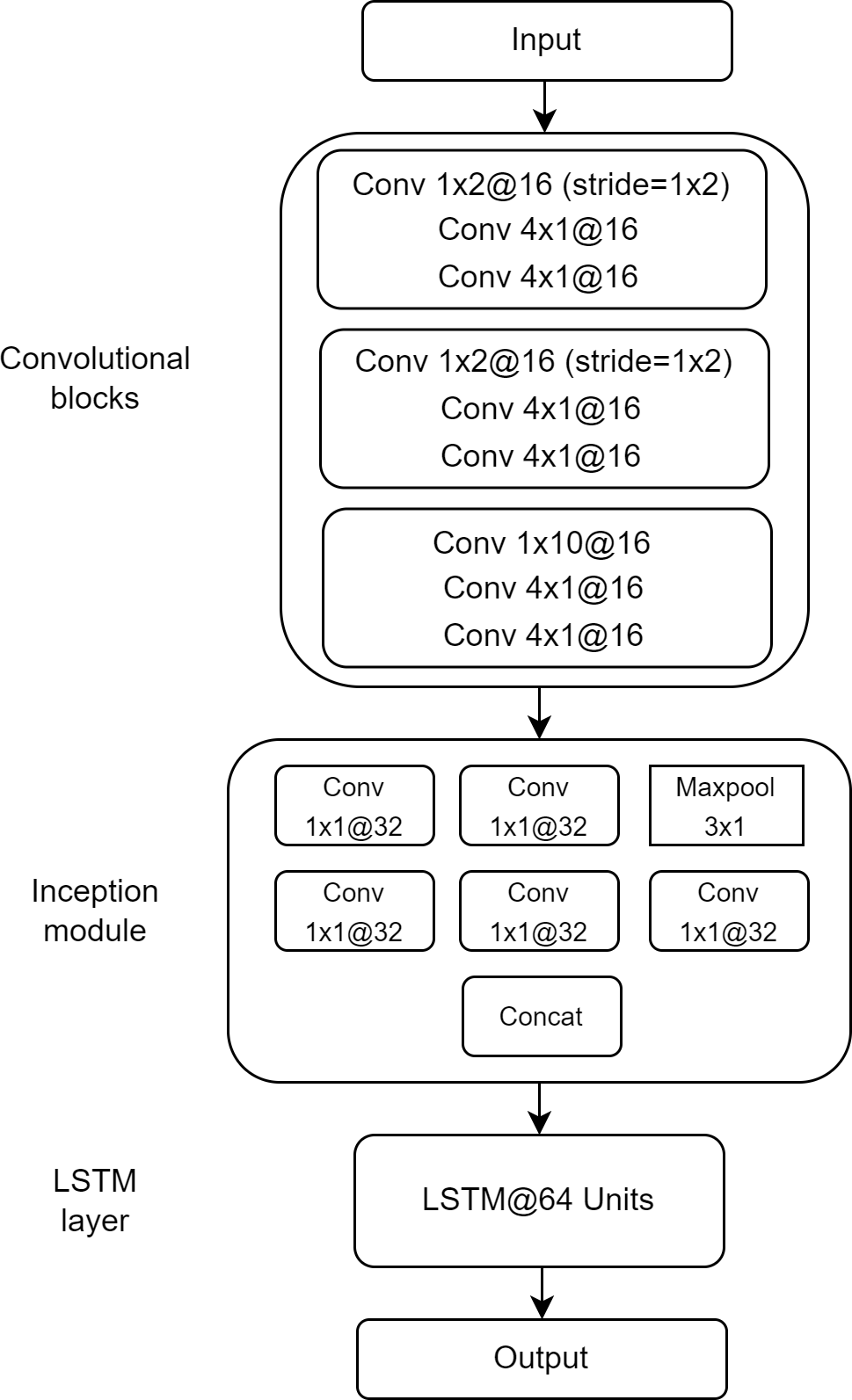}
    \caption{DeepLOB architecture sourced from Zhang et al. \cite{DeepLOB}. "Conv $1\times2$ @16" means there is 16 filters of size $1\times2$ in this convolutional layer.}
    \label{fig.deeplob}
\end{figure}

\begin{figure}[h]
    \centering
    \includegraphics[scale=1]{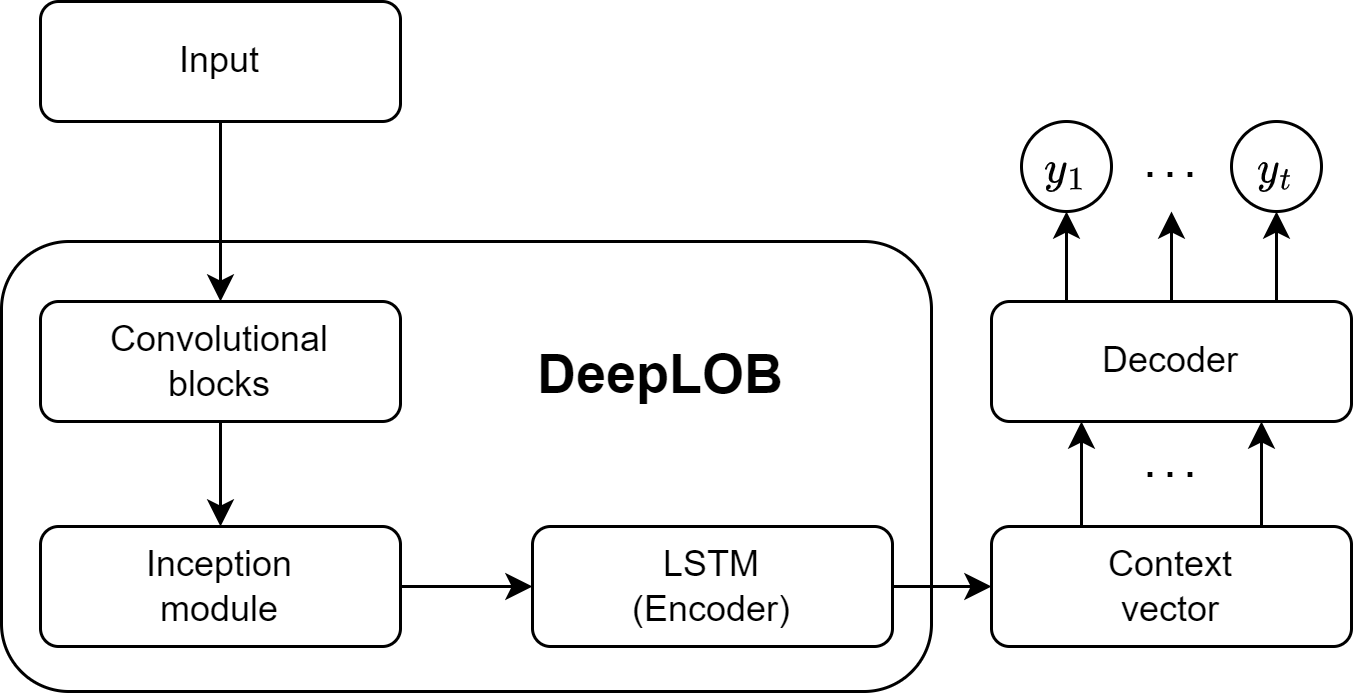}
    \caption{DeepLOB-Seq2Seq and DeepLOB-Attention architecture sourced from Zhang et al. \cite{LOBs2s}.}
    \label{fig.lobs2s}
\end{figure}

Besides the canonical LSTM, three more LSTM based-models are chosen for comparison to Transformer-based models. They are DeepLOB \cite{DeepLOB}, DeepLOB-Seq2Seq \cite{LOBs2s} and DeepLOB-Attention \cite{LOBs2s}  created by Zhang et al. The architecture of these three models are shown in Figure \ref{fig.deeplob} and \ref{fig.lobs2s}. Here the structures of these three models are briefly explained:\\\par
\textbf{DeepLOB} \cite{DeepLOB} DeepLOB’s architecture consists of three main components: Convolutional Blocks, an Inception Module and an LSTM layer. \\
\textit{A. Convolutional Blocks} The LOB inputs mentioned in Equation \ref{eq:1} are fed into the convolutional blocks that contain multiple convolutional layers, where the first and second convolutional blocks are more important than the third one. The first convolutional block has a layer with filter size of ($1\times2$) and stride of ($1\times2$). At each order book level, this layer summarise the price and volume information $\left\{p^{(i)}, v^{(i)}\right\}$. For the second convolutional block, it has a layer with the same filter size and stride as the first one, but it is a feature mapping for the micro-price defined by \cite{micro}:
\begin{equation}
    p^{\text {micro }}=I p_{i}^{a s k}+(1-I) p_{i}^{\text {bid }}
\end{equation}
\begin{equation}
    I=\frac{v_{i}^{b i d}}{v_{i}^{a s k}+v_{i}^{b i d}}
\end{equation}
$I$ is called the imbalance. Then the last convolutional block integrates the feature information from the previous two layers. The whole convolutional blocks work as a feature extractor.\\\par
\textit{B. Inception Module} the Inception Module employs the time series decomposition method. The input is decomposed by two $1\times1$ convolutions and one max-pooling layer into three lower-dimensional representations. Then these representations pass through convolution layers with 32 channels to be merged together. This decomposition method improves the prediction accuracy.\\\par

\textit{C. LSTM Layer} Finally, the extracted features are inputted into one LSTM layer to capture the underlying pattern and dependencies. The last output layer can be a SoftMax layer or a linear layer, which depends on the specific tasks.\\\par

\textbf{DeepLOB-Seq2Seq} \cite{LOBs2s} To generate multi-horizon predictions, Zhang et al. developed DeepLOB-Seq2Seq. The main idea is to feed the output of the Inception Module into a Seq2Seq architecture to do iterated multi-step (IMS) prediction. Seq2Seq \cite{seq2seq} architecture contains encoder and decoder constructed by recurrent neural network (RNN). Assume the sequence input is $X_{T}=\left(x_{1}, x_{2}, \cdots, x_{T}\right)$, the encoder will output a hidden state at each timestamp t:
\begin{equation}
    h_{t}=f\left(h_{t-1}, x_{t}\right)
\end{equation}
After obtaining the hidden states from the encoder, a context vector $c$ has to be constructed from these hidden states. The last hidden state or the mean of all hidden states can be taken as a context vector. Context vector work as a "bridge" between the encoder and decoder, where the context vector is utilized to initialize the decoder, and the hidden state output of the decoder at each timestamp is:
\begin{equation}
    d_{t}=f\left(d_{t-1}, y_{t-1}, c\right)
\end{equation}
Then the distribution of the output $y_t$ is:
\begin{equation}
    P\left(y_{t} \mid y_{<t}, c\right)=g\left(d_{t}, c\right)
\end{equation}
where the output of the decoder not only depends on the previous true value as input but is also conditioned on the context vector.\\\par
\textbf{DeepLOB-Attention} \cite{LOBs2s} With the same idea as the DeepLOB-Seq2Seq model, the difference of the DeepLOB-Attention model is changing the Seq2Seq architecture into Attention. The attention model \cite{DBLP:journals/corr/LuongPM15} constructs the context vector differently instead of using the last hidden state or the mean of all hidden states. Same as Seq2Seq model, the encoder outputs hidden state $h_t$ and decoder outputs hidden state $d_t$. The first step is to compute a similarity score between the hidden state $d_t$ of the decoder and each encoder state $h_i$, where the similarity score is usually calculated by dot product:
\begin{equation}
    s_{i}=h_{i}^{T} d_{t}
\end{equation}
And then, normalize the similarity scores to obtain a weight distribution by softmax:
\begin{equation}
    \left\{\alpha_{1}, \alpha_{2}, \ldots, \alpha_{S}\right\}=\operatorname{softmax}\left(\left\{s_{1}, s_{2}, \ldots, s_{S}\right\}\right)
\end{equation}
Finally, generate the context vector from the attention weights:
\begin{equation}
    c_{t}=\sum_{i=1}^{S} \alpha_{i} h_{i}
\end{equation}
After that, the process of producing the output $y_t$ is the same as the Seq2Seq model.\\\par

\subsection{Transformer}

\begin{figure}[h]
    \centering
    \includegraphics[scale=0.7]{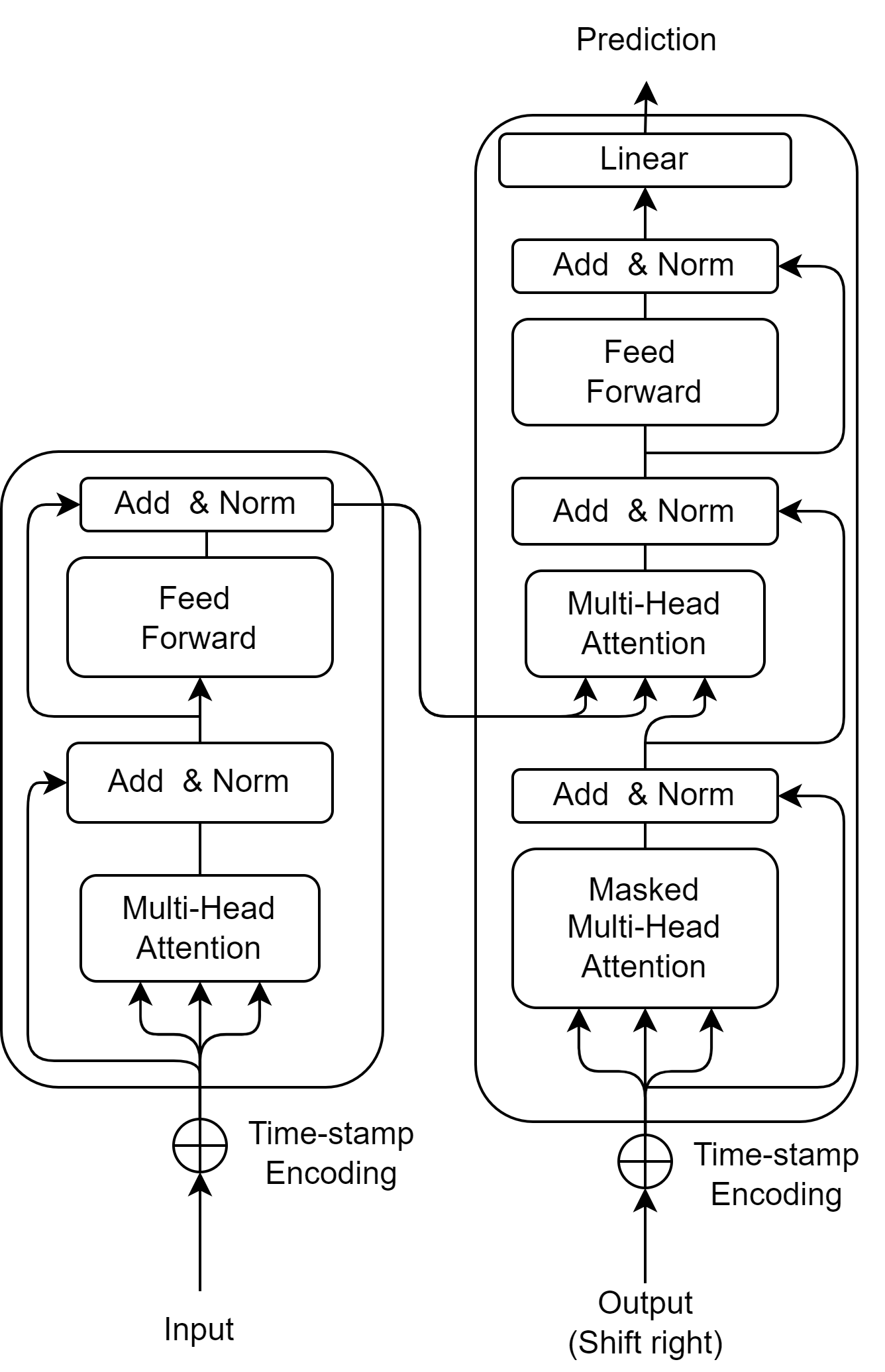}
    \caption{Transformer Structure based on Vaswani et al. \cite{attention} and Farsani et
al. \cite{TSMTSF}.}
    \label{fig.transformer}
\end{figure}

The Transformer \cite{attention} has an encoder-decoder structure that relies on the Self-attention mechanism without relying on CNN and RNN. This architecture allows the Transformer to process the long sequence and has no problem with the vanishing gradient, which means it can model the dependency regardless of the length of the input/output. A series of components forms the Transformer: Multi-head Self-attention, positional-wise feed-forward neural network, layer-normalization, and residual connection. According to Vaswani et al. \cite{attention} and Farsani et al. \cite{TSMTSF}, the architecture of the Transformer for financial time series prediction is shown in Fig.\ref{fig.transformer}.
There is a slight difference between this transformer and the vanilla one used for NLP tasks. The word embedding process is omitted, and the financial time series is fed into the transformer using time-stamp encoding. The details of time-stamp encoding will be explained in Section~\ref{tsencoding}.

\subsubsection{Multi-head Self-attention Mechanism }

Self-attention is a mechanism for finding the relevant vector in a sequence. The target of Self-attention is to compute the attention score and then extract information based on the score. According to Vaswani et al. \cite{attention}, the calculations of scaled dot-production Self-attention are as follows:
First, to compute the attention score, multiply the input vector $I$ with different learnable weight matrices $W^q$ and $W^k$ to obtain the query and key:
\begin{align}
     Q=W^qI\\
     K=W^kI
\end{align}
The next step is to calculate the attention matrix, where each element in it is an attention score:
\begin{equation}
    A=\operatorname{softmax}\left(\frac{Q K^{T}}{\sqrt{d_{k}}}\right)
\end{equation}
Where $d_k$ is the dimension of query and key.\\
Lastly, multiply the attention matrix with the values, and the final output can be obtained:
\begin{equation}
    \text { Attention }(Q, K, V)=\operatorname{softmax}\left(\frac{Q K^{T}}{\sqrt{d_{k}}}\right) V
\end{equation}
The output is the weighted sum of the relevance of different vectors in the sequence.
In the implementation of the transformer, Multi-head Self-attention is used, which is beneficial for finding different types of relevance.

\subsubsection{Learnable Time-stamp Encoding} \label{tsencoding}

\begin{figure}[h]
    \centering
    \includegraphics[scale=1]{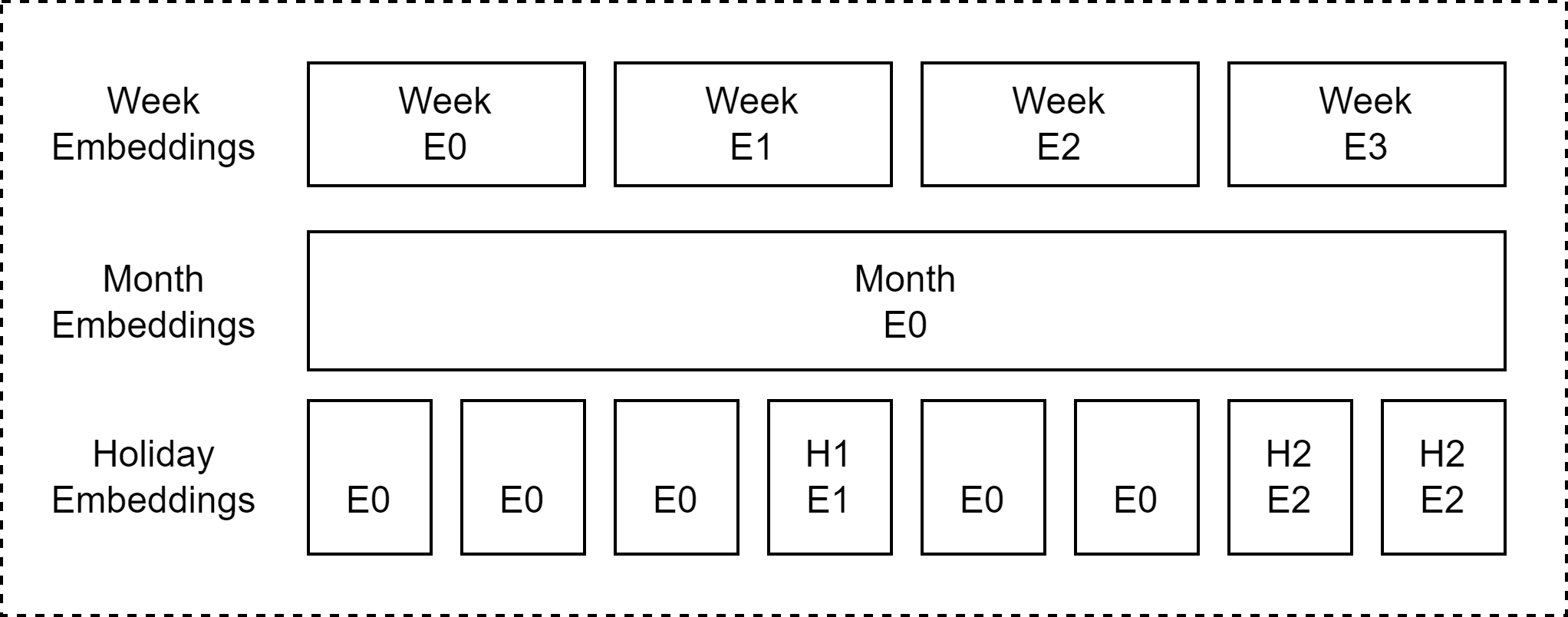}
    \caption{Time-stamp embeddings based on Zhou et al. \cite{Informer}.}
    \label{fig.timeembed}
\end{figure}

Different from the fixed sinusoid encoding used in the vanilla transformer \cite{attention}, timestamp encoding from the Informer \cite{Informer} is more informative for time series data and a similar method is applied in Autoformer \cite{Autoformer} and FEDformer \cite{FEDFormer}. The timestamp encoding method uses learnable embedding layers to produce positional information to add to the sequence, where timestamp information like a minute, hour, week, year and extra time stamps like event or holiday can be incorporated. To obtain the time-step encoding, the first step is to calculate fixed sinusoid encoding. Assume the input is $X_{t}=\left\{x_{1}, x_{2}, \ldots, x_{L_{x}} \mid x_{i} \in R^{d_{x}}\right\}_{t}$ at timestamp $t$ , where $L_x$ is the sliding window size and $d_{x}$ is the model dimensionality, then the encoding is calculated as follows:
\begin{align}
    PE_{pos, 2i}=\sin \left(\frac{p o s}{\left(2 L_{x}\right)^{\frac{2 i}{d}}}\right)\\
    PE_{pos, 2i+1}=\cos \left(\frac{p o s}{\left(2 L_{x}\right)^{\frac{2 i}{d}}}\right)
\end{align}
And then, project the original input $x_i^t$ into the model dimensionality vector $u_i^t$ using convolutional filters.
The next step is to use a learnable embedding layer $\text {SE}_{pos}$ to incorporate the timestamp information. The structure of the timestamp embedding is shown in Fig.\ref{fig.timeembed}.\\\par
Finally, add up all the calculation results above to get the final encoding:
\begin{equation}
    X_{f e e d[i]}^{t}=\alpha u_{i}^{t}+P E_{\left(L_{x} \times(t-1)+i\right)}+\sum_{p}\left[S E_{L_{x}(t-1)+i}\right]_{p}
\end{equation}
Where $i \in\left\{1, \ldots, L_{x}\right\}$ and $\alpha$ is the parameter balancing the ratio of scalar projection and embeddings.

\subsubsection{Alternative Transformer-based Models} \label{altrans}

\begin{figure}[h]
    \centering
    \includegraphics[scale=0.5]{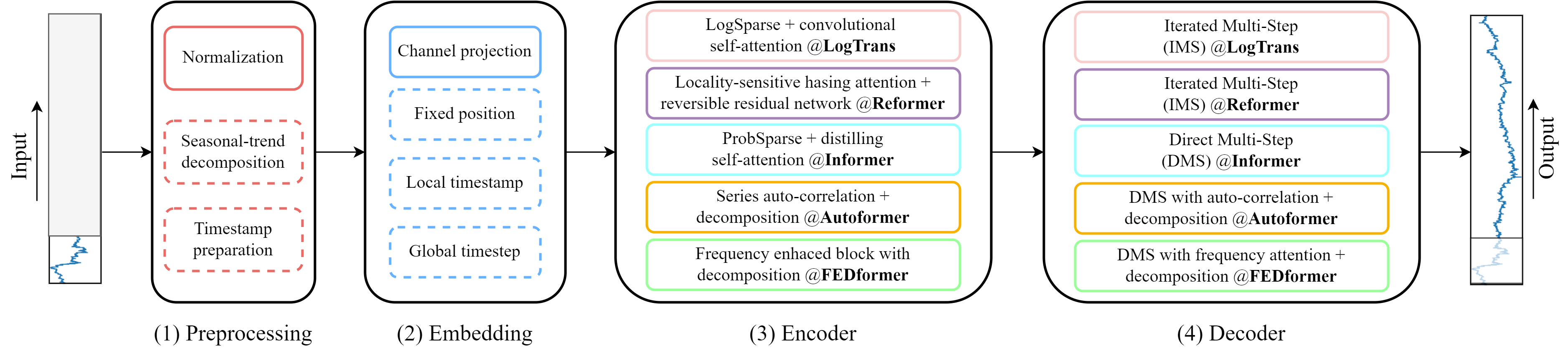}
    \caption{Architecture designs for alternative Transformer models sourced from Zeng et al.\cite{dlinear}. In step (1) and (2), the operations inside the solid box are required, while those in the dashed box can be applied optionally.}
    \label{fig.altrans}
\end{figure}

As mentioned in Section~\ref{section_2}, several new Transformer-based models \cite{Reformer, LogTrans, Informer, Autoformer, FEDFormer} are dedicated for long time series forecasting. However, they have not been tested on financial time series data. In this study, they are chosen as the alternative models to compare with the vanilla Transformer and LSTM. These models and  their relationships are briefly summarized as follows (see Figure \ref{fig.altrans}):\\\par

\textbf{LogTrans} Li et al. \cite{LogTrans} put forward LogTrans with convolutional Self-attention generating queries and keys in the Self-attention layer. This work makes the convolutional layer widely used in the attention module in the later studies \cite{Informer, Autoformer, FEDFormer}. It uses a Logsparse mask to reduce the time complexity from $O(L^2)$ to $O(L\log(L))$.\\\par
\textbf{Reformer} Reformer \cite{Reformer} changes the time complexity of the Transformer to $O(L\log(L))$ as well by using sensitive hashing instead of dot-product in the calculate of attention. It also replaces the residual connection in the vanilla Transformer with a reversible residual connection, which makes the Transformer more efficient.\\\par
\textbf{Informer} Although Reformer and LogTrans reduced the time complexity to $O(L\log(L))$, the memory consumption is still the same, so the efficiency gain is not high. Also, LogTrans and Reformer use iterated multi-step prediction (IMS), generating a single prediction in each timestamp and using that prediction iteratively to obtain multi-step prediction \cite{IMS}, which suffers from error accumulation. Informer \cite{Informer} proposed by Zhou et al. employs a ProbSparse Self-attention mechanism to achieve $O(L\log(L))$ time and memory complexity. They propose an innovative generative style decoder to make direct multi-step(DMS) prediction, which is to generate the multi-step prediction at once in one forward procedure \cite{DMS}. This method speed up the long-term forecast compared to the LogTrans and Reformer. The DMS forecast method and learnable timestamp encoding are applied in the Autoformer \cite{Autoformer} and FEDformer \cite{FEDFormer}.\\\par
\textbf{Autoformer} The optimization of Transformer on time series prediction is a trade-off between efficiency and information utilization. Depending on the structure of Informer, Autoformer \cite{Autoformer} reduced the time complexity with an Auto-Correlation mechanism rather than making Self-attention sparse in LogTrans and Informer, which can preserve the information well and measure the sub-series dependency. Time series decomposition is a method commonly used in time series analysis to deconstruct the time series into several components \cite{STL,TS}. The underlying temporal pattern can be revealed from these components to make the time series more predictable \cite{alma991000567577201591}. Autoformer first embeds the time series decomposition as an inner neural block to derive the trend-cyclical component from the input sequence and the seasonal component from the difference between the trend-cyclical component and the input sequence. This new decomposition architecture can deconstruct time series to use the series periodicity to update attention.\\\par
\textbf{FEDformer} Based on the decomposition architecture of Autoformer, Zhou et al. \cite{FEDFormer} builds the FEDformer to use Fourier transform and Wavelet transform, which are in the frequency domain as a new decomposition method to reach linear complexity $O(L)$. .\\\par


\section{Experimentation Result and Evaluation}\label{section_5}

\subsection{Comparison of LOB Mid-Price Prediction}

The first task to compare transformer versus LSTM is the mid-price prediction. In this task, predicting the absolute value of the mid-price is similar to the previous work's \cite{LogTrans, Informer, Autoformer, FEDFormer,liu2022pyraformer} experiment on non-financial datasets.

\subsubsection{Experiment Setting for LOB Mid-Price Prediction}

\textbf{Dataset} All the experiments are based on cryptocurrency LOB data, which are collected in real-time from Binance Exchange using cryptofeed \cite{cryptofeed} WebSocket API and saved to the database using kdb+tick triplet \cite{novotny2019machine}. In this experiment, one-day LOB data of product BTC-USDT (Bitcoin-U.S. dollar tether) on $2022.07.15$. containing $863397$ ticks is utilized. The time interval between each ticks is not evenly spaced. The time interval is $0.1$ second on average. The first $70\%$ data is used to construct the training set, and the rest $10\%$ and $20\%$ of data are used for validation and testing. The reason why only using one day of LOB data is that it is enough to train for a task prediction for absolute value without overfitting according to previous works’ experiments on non-financial datasets.\\
\textbf{Models} For the comparison purpose, I choose canonical LSTM and vanilla Transformers along with four Transformer-based models: FEDformer \cite{FEDFormer}, Autoformer \cite{Autoformer}, Informer \cite{Informer} and Reformer \cite{Reformer}. For the implementation of Transformer-based models, they are taken from open-source code repositories \cite{Autoformer_repo, FEDformer_repo}. The implementation of vanilla Transformer \cite{attention}, Reformer \cite{Reformer}, Informer \cite{Informer} and Autoformer \cite{Autoformer} are from the Autoformer repository \cite{Autoformer_repo}. The implementation of FEDformer \cite{FEDFormer} is from its own repository \cite{FEDformer_repo}. \\
\textbf{Training setting} The dataset is normalized by the z-score normalization method. The validation set and the test set are normalized by the mean and standard deviation of the training set.  All the models are trained for 10 epochs using the Adaptive Momentum Estimation optimizer and L2 loss with early stopping. The batch size is 32, and the initial learning rate is 1e-4. All models are implemented by Pytorch \cite{pytorch} and trained on a single NVIDIA RTX A5000 GPU with 24 GB memory.

\subsubsection{Result and Analysis for LOB Mid-Price Prediction}

\begin{table}[hbt!]
\centering
\resizebox{\columnwidth}{!}{%
\begin{tabular}{@{}c|lccccccccccc@{}}
\toprule
Models  & \multicolumn{2}{c|}{FEDformer}           & \multicolumn{2}{c|}{Autoformer} & \multicolumn{2}{c|}{Informer} & \multicolumn{2}{c|}{Reformer} & \multicolumn{2}{c|}{Transformer} & \multicolumn{2}{c}{LSTM} \\ \midrule
Metrics & \multicolumn{1}{c}{MSE} & MAE            & MSE             & MAE           & MSE           & MAE           & MSE           & MAE           & MSE             & MAE            & MSE         & MAE        \\ \midrule
96      & \textbf{0.0793}         & \textbf{0.179} & $0.0926$        & $0.201$       & $1.411$       & $0.543$       & $2.186$       & $0.619$       & $2.836$         & $0.696$        & $0.104$     & $0.204$    \\
192     & \textbf{0.155}          & \textbf{0.257} & $0.176$         & $0.279$       & $1.782$       & $0.749$       & $1.842$       & $0.824$       & $2.799$         & $0.832$        & $0.195$     & $0.287$    \\
336     & \textbf{0.274}          & \textbf{0.348} & $0.319$         & $0.376$       & $2.080$       & $0.830$       & $9.218$       & $1.947$       & $1.456$         & $0.665$        & $0.315$     & $0.369$    \\
720     & \textbf{0.608}          & \textbf{0.514} & $0.643$         & $0.539$       & $2.808$       & $1.093$       & $72.57$       & $6.824$       & $4.306$         & $1.297$        & $0.771$     & $0.587$    \\ \bottomrule
\end{tabular}%
}
\caption{Mid price prediction result with different prediction lengths $k\in\{96,192,336,720\}$ in test set. The input window size is set to $96$ (MSE's unit is in $10^{-2}$ and MAE's unit is in $10^{-1}$).}
\label{table:1}
\end{table}
\textbf{Quantitative result} To evaluate the performance of different models, following the previous works \cite{LogTrans, Informer, Autoformer, FEDFormer,liu2022pyraformer}, the performance metrics consist of Mean Square Error (MSE) and Mean Absolute Error (MAE), representing the prediction error. Lower MSE or MAE indicates the model has less prediction error. MSE and MAE are calculated by:
\begin{equation}
    M S E=\frac{1}{n} \sum_{i=1}^{n}\left(Y_{i}-\widehat{Y}_{i}\right)^{2}
\end{equation}
\begin{equation}
    M A E=\frac{1}{n} \sum_{i=1}^{n}\left|Y_{i}-\hat{Y}_{i}\right|
\end{equation}
where $Y_{i}$ is the true value and $\widehat{Y}_{i}$ is the predicted value. $n$ is the number of ticks. \\
The results of all models are shown in Table \ref{table:1}. From the table, these outcomes can be summarized:\\
1. Both FEDformer and Autoformer outperform LSTM and FEDformer has the best performance in all the prediction lengths. FEDformer and Autoformer give a large increase in performance in terms of MSE and MAE compared to LSTM. For FEDformer, it gives $24\% (0.104 \rightarrow 0.0793)$ MSE reduction on 96 prediction length and $21\%(0.771 \rightarrow 0.608)$ on 336 prediction length. For Autoformer, it gives $11\%(0.104 \rightarrow 0.0926)$ MSE reduction on 96 prediction length and $16\%(0.771 \rightarrow 0.643)$ MSE reduction on 336 prediction length. These results show that both Autoformer and FEDformer perform well in terms of MSE and MAE because of their low error and long-term robustness. \\
2. Although FEDformer and Autoformer's MSE and MAE are low on this task, LSTM is relatively not bad on mid-price prediction. LSTM outperforms the other three models: Informer, Reformer, and vanilla Transformer, which indicates that LSTM is robust in handling LOB data, while transformer-based models require lots of modification to perform well. \\
3. Vanilla Transformer model has worse performance on prediction lengths $96$ and $192$ and Reformer has worse performance on prediction lengths $336$ and $720$ because they suffered from error accumulation during the IMS prediction process. Informer’s worse performance than LSTM is mainly due to its sparse version of attention, leading to information loss on the time series. 
\begin{figure}[h]
    \centering
    \includegraphics[scale=0.08]{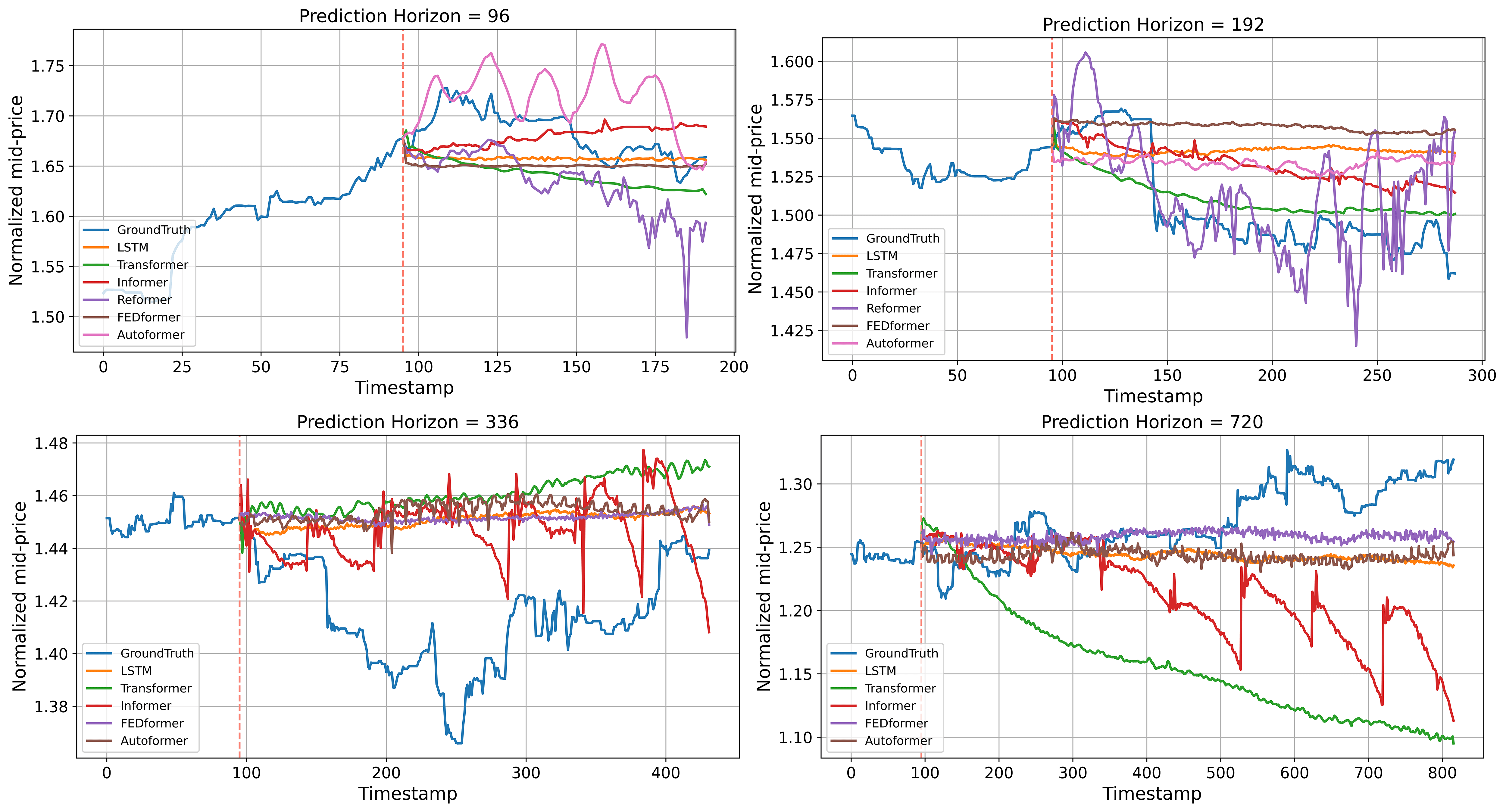}
    \caption{Illustration of normalized forecasting outputs with 96 input window size and $\{96,192,336,720\}$ prediction lengths. Each timestamp is one tick. Reformer's results are not plotted in the lower two panels for better visualisation.}
    \label{fig.reg_all}
\end{figure} \\
\textbf{Qualitative results} The prediction results of compared models on all the prediction horizons are shown in Figure \ref{fig.reg_all}. When the prediction horizon is $96$, Autoformer and Reformer are able to generate a proper trend for the future mid-price, while other models generate almost a flat line as predictions. On the prediction horizon of $192$, almost all the models’ predictions plateau except the Reformer, but Reformer’s result becomes more stochastic than the prediction horizon of $96$. For larger prediction horizons $336$ and $720$, all the models can hardly predict a proper trend and Reformer’s result in not plotted because it becomes too stochastic. \\\par
Based on the qualitative results above, although Autoformer and FEDformer outperform LSTM in terms of MSE and MAE, their actual prediction performance is far more inadequate for high-frequency trading. The analysis from Figure \ref{fig.reg_all} is just “eyeballing” whether the model is good. In this case, another formal metric out of sample $R^2$ is added here to judge the prediction quality. According to Lewis-Beck \cite{r_square}, $R^2$ determine how well the prediction result $Y$ can be explained by the input $X$, and higher $R^2$ indicates the model has a better fit for the predicted value. Out of sample $R^2$ is defined by:
\begin{equation}
    R^{2}=1-\frac{\sum_{i=1}^{n}\left(Y_{i}-\widehat{Y}_{i}\right)^{2}}{\sum_{i=1}^{n}\left(Y_{i}-\bar{Y}\right)}
\end{equation}
where $\bar{Y}=\frac{1}{n} \sum_{i=1}^{n} Y_{i}$; $\widehat{Y}_{i}$ is the predicted value and $Y_{i}$ is the ground truth.\\
Here the out-of-sample $R^2$ is calculated based on the price difference. Please note that the price difference here differs from the one mentioned in Section~\ref{task2}. The price difference here is calculated from the absolute price prediction result, while the one in Section~\ref{task2} is the direct prediction target. The result of $R^2$ is shown in Table \ref{table.reg_r}.
\begin{table}[]
\centering
\resizebox{\textwidth}{!}{%
\begin{tabular}{@{}c|cccccc@{}}
\toprule
Models & Autoformer & FEDformer & Informer & Reformer & LSTM   & Transformer \\ \midrule
96     & -0.753     & -0.237    & -43.811  & -69.080  & -0.946 & -87.899     \\
192    & -0.596     & -0.205    & -25.281  & -26.792  & -0.644 & -43.368     \\
336    & -1.032     & -0.364    & -20.123  & -63.252  & -0.414 & -13.035     \\
720    & -0.521     & -0.189    & -7.760   & -137.322 & -0.589 & -16.314     \\ \bottomrule
\end{tabular}%
}
\caption{Average of out of sample $R^2$ result with different prediction lengths $k\in\{96,192,336,720\}$.}
\label{table.reg_r}
\end{table}\\
From the table, all the out-of-sample $R^2$ values are negative for all models. However, according to Lewis-Beck \cite{r_square}, $R^2$ at least needs to be larger than zero to indicate that input $X$ can explain output $Y$. This indicates that predictions for absolute mid-price are useless for trading purposes. The metrics of MSE and MAE obscure the real quality of prediction results, highlighting the importance of $R^2$ calculated based on the price difference. \\\par

To sum up, although Autoformer and FEDformer have lower MSE and MAE than LSTM, their prediction result is not helpful and practical for trading. The more sensible way is to directly use the price difference as the prediction target. 

\subsection{Comparison of LOB Mid-Price Diff Prediction}

Both transformer-based models and LSTM are unable to generate the satisfactory result on the mid-price prediction, so this task turns to the mid-price difference prediction. The mid-price difference is a useful alpha-term structure in trading for traders and market makers \cite{Kolm2021DeepOF}. 

\subsubsection{Experiment Setting for LOB Mid-Price Diff Prediction}

\textbf{Dataset} The dataset for this experiment is collected the same way as the last experiment, but the dataset size becomes larger to avoid overfitting. In this experiment, four days of LOB data for the product BTC-USDT from 2022.07.03 (inclusive) to 2022.07.06 (inclusive) is used, containing 3432211 ticks. The first $80\%$ of data is used as a training set, and the rest $20\%$ is split in half for validation and testing. \\
\textbf{Models and Limitations} Five models are being compared in this experiment: canonical LSTM \cite{lstm}, vanilla transformer \cite{attention}, CNN-LSTM (DeepLOB \cite{DeepLOB} model used for regression), Informer \cite{Informer} and Reformer \cite{Reformer}. State-of-the-art FEDformer and Autoformer are not compared in this task because their time decomposition structure is limited in handling price difference series. As mentioned in Section~\ref{task3}, the price difference is approximately stationary. The time decomposition method is useful for the non-stationary time series, such as the mid-price series, where it can extract meaningful patterns. In contrast, the price difference series is approximately stationary, and little meaningful information can be extracted from it. Therefore, FEDformer and Autoformer can only produce a poor result in this task and will not be compared below. \\
\textbf{Training settings} the training setting is the same as the last experiment.

\subsubsection{Result and analysis for LOB Mid-Price Diff Prediction}

\begin{figure}[h]
    \centering
    \includegraphics[scale=0.5]{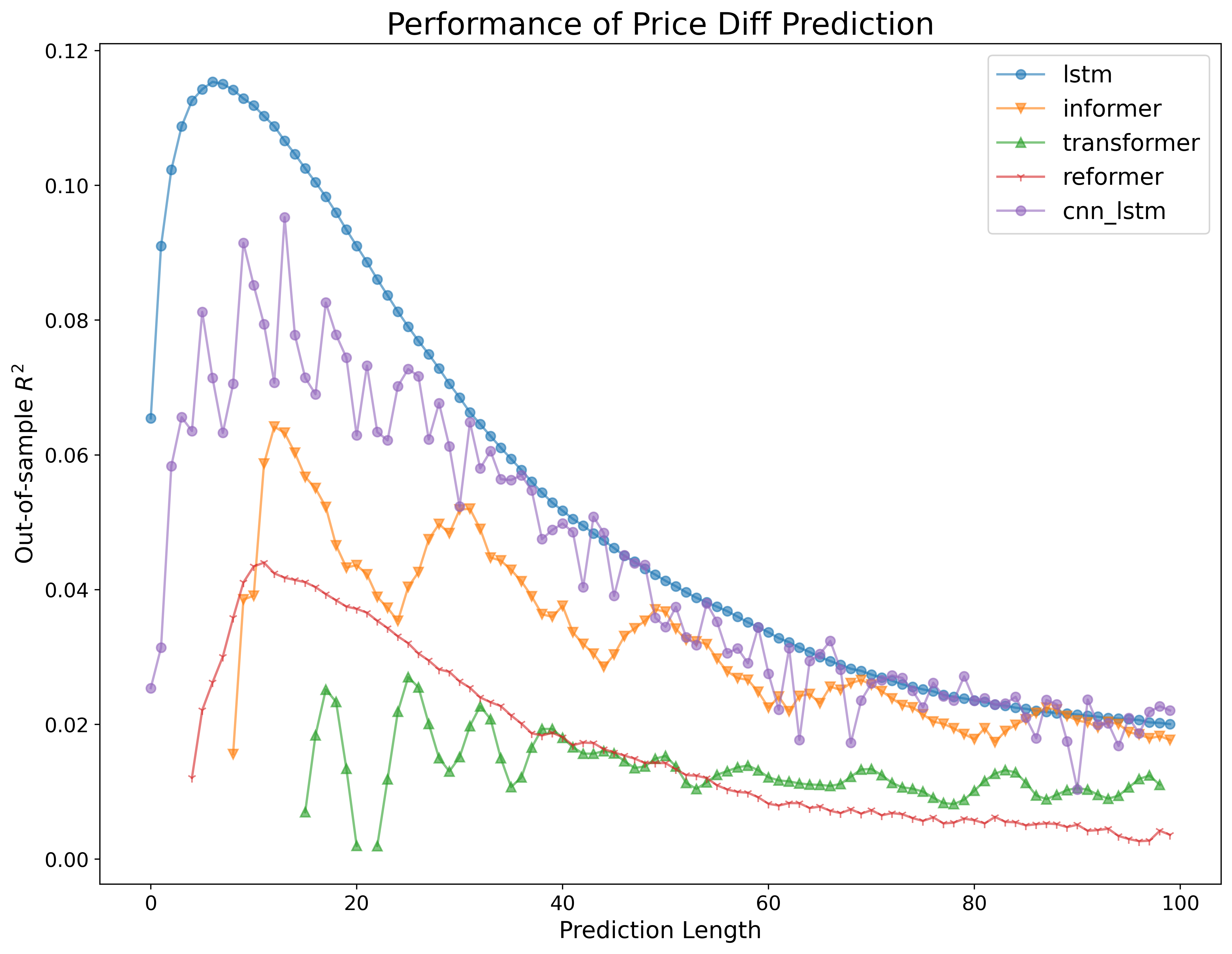}
    \caption{Performance of price difference prediction with input window size 100 and prediction length 100. Negative data points are not plotted for ease of visualization.}
    \label{fig.price_diff}
\end{figure}
Following the previous works \cite{Kolm2021DeepOF}, out of sample $R^2$ is the evaluation metric for this task. The performance of all the models is shown in Figure \ref{fig.price_diff}. The canonical LSTM achieves the best performance among all models, which reaches the highest $R^2$ around $11.5\%$ in forecast length $5$ to $15$. For CNN-LSTM, it has comparable performance to LSTM. On the other hand, Informer, Reformer and Transformer have worse $R^2$ than LSTM, but their $ R^2$ trend is similar. In short, for the price difference prediction task, LSTM-based models is more stable and more robust than Transformer-based models. This result is in expectation because Reformer, Informer and Transformer already have worse performance than LSTM in mid-price prediction task because of their shortcomings. At the same time, the state-of-art FEDformer and Autoformer cannot be applied because of their limitation. In order to let these state-of-the-art transformer-based models make a meaningful prediction, a new structure is designed in the next part, and it is applied to the price movement prediction task.

\subsection{Comparison of LOB Mid-Price Movement Prediction}\label{task3_res}

\subsubsection{Innovative Architecture on Transformer-based Methods} \label{innov_1}

\begin{figure}[h]
    \centering
    \includegraphics[scale=0.8]{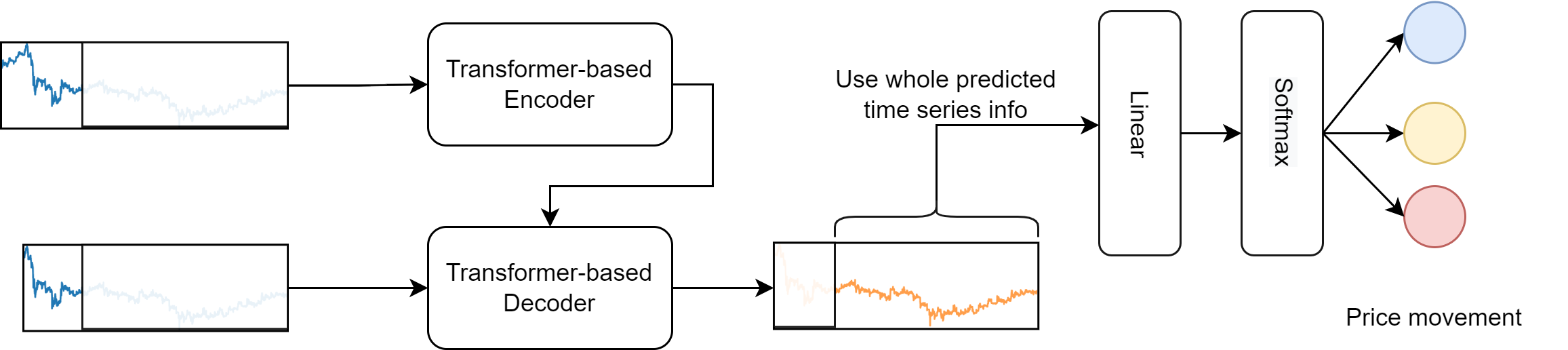}
    \caption{New architecture of transformer-based model for LOB mid-price movement prediction.}
    \label{fig.trans_innov}
\end{figure}

The alternative Transformer-based models mentioned in Section~\ref{altrans} mainly focus on the long time series forecasting problem, which is a regression task. For the LOB data, it is easy to adapt these models for the mid-price prediction and mid-price difference prediction because both are regression problems. For the mid-price movement prediction task, the model needs to produce a classification result for the future, and there are few existing Transformer models specialized available for this task. In contrast, most of the Transformer-based models are designed for classification tasks without forecasting, such as sentiment analysis, spam detection and pos-tagging in NLP. In this case, adapting the existing Transformer-based models to do price movement forecasting is necessary. I first adapt Transformer-based models in price movement forecasting and want to facilitate transformer development in this specific task. The new architecture of the transformer-based model is shown in Figure \ref{fig.trans_innov}. The details are explained as follows:\\
Predicting the next mid-price movement based on the past price and volume information is an one step ahead of prediction.
A straightforward method to adapt the transformer-based model is to pass the next predicted mid-price into a softmax activation. However, this method performs poorly because it only considers the past mid-price information and ignores future ones. It is worth noting that in the labelling process in Section~\ref{task3}, previous and next $k$ mid-price information are utilized. In this case, I adapt the existing transformer-based models to feed the whole predicted mid-price sequence into a linear layer and finally pass through a softmax activation function to generate price movement output. This adaptation will benefit those transformer-based models using the DMS forecasting method because they have fewer errors in the long-time series prediction process.

\subsubsection{DLSTM: Innovation on LSTM-based Methods}

Inspired by the Dlinear model \cite{dlinear} and Autoformer, combining the merits of time decomposition with the LSTM, a new model  named DLSTM is designed.
DLSTM is designed based on these three observations: Firstly, the time series decomposition method is capable of increasing the performance, especially embedding this process by neural blocks in previous works \cite{DeepLOB, Autoformer, FEDFormer}. Secondly, LSTM is a robust and simple model for multiple forecasting tasks. Thirdly, Dlinear beats other Transformer-based models in some long time series forecasting tasks thanks to the time series decomposition method and DMS prediction. However, predicting price movement is one step ahead prediction, where the model will not suffer from the error accumulation effect. In this case, it is sensible to replace linear with LSTM, because LSTM is a model well-known better than linear in handing time series.
\begin{figure}[h]
    \centering
    \includegraphics[scale=0.75]{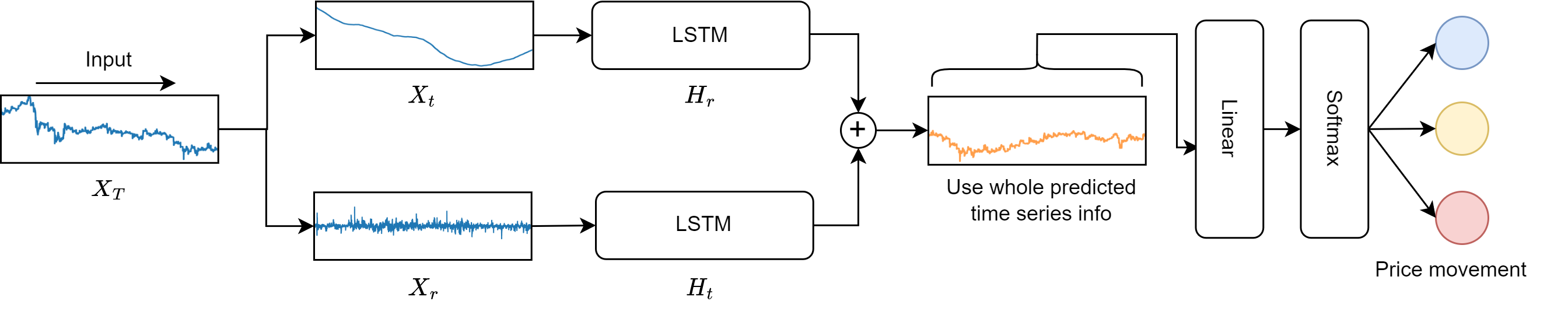}
    \caption{Architecture of DLSTM}
    \label{fig.dlstm}
\end{figure}\\
The architecture of DLSTM is shown in Figure \ref{fig.dlstm}. The main difference between Dlinear and DLSTM is that the LSTM layers replace the Linear layer. According to the time decomposition method introduced in Autoformer \cite{Autoformer}, in the prediction process, assume there is a time series $X_{T}=\left(x_{1}, x_{2}, \ldots, x_{T}\right)$, first decompose it into Trend series by the moving average:
\begin{equation}
    X_{t} = AvgPool(Padding(X_{T}))
\end{equation}
where $AvgPool(\cdot)$ is the average pooling operation and the $Padding(\cdot)$ is to fix the input length.\\
And then the Remainder series is calculated by $X_{r}=X_{T}-X_{t}$. After that, these two series are inputted into two LSTM layers. Finally, the hidden states $H_{t}$ and $H_{r}$ produced by two LSTM layers will be added together and then pass through a linear and softmax activation to generate the final price movement result.

\subsubsection{Setting for LOB Mid-Price Movement Prediction}
\label{task3_setting}

\textbf{Dataset} In this experiment, the largest dataset among three tasks is utilized to avoid over-fitting and test the model’s robustness. The whole dataset contains 12 days of LOB data of product ETH-USDT (Ethereum-U.S. dollar tether) from 2022.07.03 (inclusive) to 2022.07.14 (inclusive), containing 10255144 ticks. The training and testing data are taken from the first six days and the last three days, and the left data are used for validation. The test set is also used for the simple trading simulation. \\
\textbf{Models} Thanks to the innovative structure mentioned in Section~\ref{innov_1}, most of the transformer-based models can be adapted and applied in this task for comparison, which are: Vanilla Transformer \cite{attention}, Reformer \cite{Reformer}, Informer \cite{Informer}, Autoformer \cite{Autoformer}, FEDformer \cite{FEDFormer}. On the other hand, all the LSTM-based models are compared in this task as well, which are: canonical LSTM \cite{lstm}, DLSTM, DeepLOB \cite{DeepLOB}, DeepLOB-Seq2Seq \cite{LOBs2s}, DeepLOB-Attention \cite{LOBs2s}. Besides these models, a simple MLP model is built as a baseline. The implementation of the Transformer-based models are based on the code repository mentioned above. The implementation of DeepLOB \cite{DeepLOB}, DeepLOB-Seq2Seq \cite{LOBs2s}, DeepLOB-Attention \cite{LOBs2s} are based on two repositories \cite{DeepLOB_repo, s2sLOB_repo}. For the DLSTM, it is inspired by code of Dlinear\cite{dlinear} model from its repository \cite{dlinear_repo}.  \\
\textbf{Training settings} The batch size for training is set to 64 and the loss function is changed to Crossentropy loss. Other training settings are the same as the last experiment.

\subsubsection{Result and analysis for LOB Mid-Price Movement Prediction}
The performance of models is evaluated by classification metrics: accuracy and the mean of precision, recall and F1-score. Result are shown in Table \ref{table:class1} and Table \ref{table:class2}. 
\begin{table}[]
\centering
\begin{tabular}{@{}lllll@{}}
\toprule
\multicolumn{1}{l|}{\textbf{Model}}    & \textbf{Accuracy}                     & \textbf{Precision}                    & \textbf{Recall}                       & \textbf{F1}                           \\ \midrule
\multicolumn{5}{c}{\textbf{Prediction Horizon k = 20}}                                                                                                                                                 \\ \midrule
\multicolumn{1}{l|}{MLP}               & 61.58                                 & 61.70                                 & 61.58                                 & 61.47                                 \\
\multicolumn{1}{l|}{LSTM}              & 62.77                                 & 62.91                                 & 62.77                                 & 62.78                                 \\
\multicolumn{1}{l|}{DeepLOB}           & 70.29                                 & 70.58                                 & 70.30                                 & 70.24                                 \\
\multicolumn{1}{l|}{DeepLOB-Seq2Seq}   & {\color[HTML]{006494} {\ul 70.40}}    & {\color[HTML]{006494} {\ul 70.79}}    & {\color[HTML]{006494} {\ul 70.42}}    & {\color[HTML]{006494} {\ul 70.37}}    \\
\multicolumn{1}{l|}{DeepLOB-Attention} & 70.04                                 & 70.26                                 & 70.03                                 & 70.01                                 \\
\multicolumn{1}{l|}{Autoformer}        & 68.89                                 & 68.99                                 & 68.89                                 & 68.91                                 \\
\multicolumn{1}{l|}{FEDformer}         & 65.37                                 & 65.70                                 & 65.37                                 & 65.20                                 \\
\multicolumn{1}{l|}{Informer}          & 68.71                                 & 68.82                                 & 68.72                                 & 68.71                                 \\
\multicolumn{1}{l|}{Reformer}          & 68.01                                 & 68.26                                 & 68.00                                 & 67.95                                 \\
\multicolumn{1}{l|}{Transformer}       & 67.80                                 & 67.99                                 & 67.81                                 & 67.77                                 \\
\multicolumn{1}{l|}{DLSTM}             & {\color[HTML]{B22222} \textbf{73.10}} & {\color[HTML]{B22222} \textbf{74.01}} & {\color[HTML]{B22222} \textbf{73.11}} & {\color[HTML]{B22222} \textbf{73.11}} \\ \midrule
\multicolumn{5}{c}{\textbf{Prediction Horizon k = 30}}                                                                                                                                                 \\ \midrule
\multicolumn{1}{l|}{MLP}               & 59.19                                 & 59.30                                 & 58.70                                 & 58.48                                 \\
\multicolumn{1}{l|}{LSTM}              & 60.64                                 & 60.47                                 & 60.45                                 & 60.45                                 \\
\multicolumn{1}{l|}{DeepLOB}           & 67.23                                 & 67.26                                 & 67.17                                 & 67.15                                 \\
\multicolumn{1}{l|}{DeepLOB-Seq2Seq}   & 67.56                                 & 67.73                                 & 67.53                                 & 67.49                                 \\
\multicolumn{1}{l|}{DeepLOB-Attention} & 67.21                                 & 67.39                                 & 66.98                                 & 66.96                                 \\
\multicolumn{1}{l|}{Autoformer}        & {\color[HTML]{006494} {\ul 67.93}}    & {\color[HTML]{006494} {\ul 67.86}}    & {\color[HTML]{006494} {\ul 67.77}}    & {\color[HTML]{006494} {\ul 67.77}}    \\
\multicolumn{1}{l|}{FEDformer}         & 66.57                                 & 66.44                                 & 66.05                                 & 65.83                                 \\
\multicolumn{1}{l|}{Informer}          & 65.41                                 & 65.33                                 & 65.14                                 & 65.13                                 \\
\multicolumn{1}{l|}{Reformer}          & 64.28                                 & 64.31                                 & 64.08                                 & 64.06                                 \\
\multicolumn{1}{l|}{Transformer}       & 64.25                                 & 64.16                                 & 64.13                                 & 64.13                                 \\
\multicolumn{1}{l|}{DLSTM}             & {\color[HTML]{B22222} \textbf{70.61}} & {\color[HTML]{B22222} \textbf{70.83}} & {\color[HTML]{B22222} \textbf{70.63}} & {\color[HTML]{B22222} \textbf{70.59}} \\ \bottomrule
\end{tabular}
\caption{Experiment results of Mid Price Movement for prediction horizons 20 and 30. {\color[HTML]{B22222} \textbf{Red Bold}} represents the best result and {\color[HTML]{006494} {\ul blue underline}} represents the second best result.}
\label{table:class1}
\end{table}

\begin{table}[]
\centering
\begin{tabular}{@{}lllll@{}}
\toprule
\multicolumn{1}{l|}{Model}             & Accuracy                              & Precision                             & Recall                                & F1                                    \\ \midrule
\multicolumn{5}{c}{Prediction Horizon k = 50}                                                                                                                                                          \\ \midrule
\multicolumn{1}{l|}{MLP}               & 55.65                                 & 55.71                                 & 55.62                                 & 54.98                                 \\
\multicolumn{1}{l|}{LSTM}              & 62.77                                 & 62.91                                 & 62.77                                 & 62.78                                 \\
\multicolumn{1}{l|}{DeepLOB}           & 63.32                                 & 63.69                                 & 63.32                                 & 63.37                                 \\
\multicolumn{1}{l|}{DeepLOB-Seq2Seq}   & 63.62                                 & 64.04                                 & 63.61                                 & 63.59                                 \\
\multicolumn{1}{l|}{DeepLOB-Attention} & {\color[HTML]{006494} {\ul 64.05}}    & {\color[HTML]{006494} {\ul 64.19}}    & {\color[HTML]{006494} {\ul 64.04}}    & {\color[HTML]{006494} {\ul 63.94}}    \\
\multicolumn{1}{l|}{Autoformer}        & 60.17                                 & 60.64                                 & 60.12                                 & 58.40                                 \\
\multicolumn{1}{l|}{FEDformer}         & 63.46                                 & 63.44                                 & 63.42                                 & 62.52                                 \\
\multicolumn{1}{l|}{Informer}          & 61.76                                 & 61.64                                 & 61.74                                 & 61.55                                 \\
\multicolumn{1}{l|}{Reformer}          & 60.43                                 & 60.79                                 & 60.42                                 & 60.37                                 \\
\multicolumn{1}{l|}{Transformer}       & 59.51                                 & 59.78                                 & 59.51                                 & 59.46                                 \\
\multicolumn{1}{l|}{DLSTM}             & {\color[HTML]{B22222} \textbf{67.45}} & {\color[HTML]{B22222} \textbf{67.96}} & {\color[HTML]{B22222} \textbf{67.45}} & {\color[HTML]{B22222} \textbf{67.59}} \\ \midrule
\multicolumn{5}{c}{Prediction Horizon k = 100}                                                                                                                                                         \\ \midrule
\multicolumn{1}{l|}{MLP}               & 57.03                                 & 56.03                                 & 56.36                                 & 56.01                                 \\
\multicolumn{1}{l|}{LSTM}              & 53.49                                 & 52.83                                 & 52.82                                 & 52.36                                 \\
\multicolumn{1}{l|}{DeepLOB}           & 58.12                                 & 58.50                                 & 57.92                                 & 57.86                                 \\
\multicolumn{1}{l|}{DeepLOB-Seq2Seq}   & 58.30                                 & 58.43                                 & 57.93                                 & 57.77                                 \\
\multicolumn{1}{l|}{DeepLOB-Attention} & 59.16                                 & {\color[HTML]{006494} {\ul 58.59}}    & {\color[HTML]{006494} {\ul 58.65}}    & {\color[HTML]{006494} {\ul 58.50}}    \\
\multicolumn{1}{l|}{Autoformer}        & {\color[HTML]{006494} {\ul 59.18}}    & {\color[HTML]{000000} 58.34}          & 58.40                                 & 57.83                                 \\
\multicolumn{1}{l|}{FEDformer}         & 57.97                                 & 56.97                                 & 56.62                                 & 54.14                                 \\
\multicolumn{1}{l|}{Informer}          & 56.11                                 & 56.15                                 & 55.85                                 & 55.81                                 \\
\multicolumn{1}{l|}{Reformer}          & 54.92                                 & 54.47                                 & 54.53                                 & 54.47                                 \\
\multicolumn{1}{l|}{Transformer}       & 55.42                                 & 55.04                                 & 54.92                                 & 54.72                                 \\
\multicolumn{1}{l|}{DLSTM}             & {\color[HTML]{B22222} \textbf{63.73}} & {\color[HTML]{B22222} \textbf{63.02}} & {\color[HTML]{B22222} \textbf{63.18}} & {\color[HTML]{B22222} \textbf{63.05}} \\ \bottomrule
\end{tabular}
\caption{Experiment results of Mid Price Movement for prediction horizons 50 and 100.{\color[HTML]{B22222} \textbf{Red Bold}} represents the best result and {\color[HTML]{006494} {\ul blue underline}} represents the second best result.}
\label{table:class2}
\end{table}
A few outcomes can be observed from the result: \\
1. DLSTM outperforms all the previous LSTM-based and Transformer-based models. It achieves the highest accuracy, precision, recall and F1 score in all the prediction horizons. This result shows that the time series decomposition structure originating from Autoformer can effectively handle time series, especially when combined with a simple LSTM model. DLSTM is making one step ahead prediction for the mid-price movement, so it will not suffer from error accumulation from the DMS prediction process. \\
2. DeepLOB-Attention model has the second best result in horizon 50 and 100 (excluding accuracy). DeepLOB-Seq2Seq has the second best result for prediction horizon 20. This result indicates that the encode-decoder structure and attention mechanism can contribute to the prediction performance because the autoregressive process can correlate the mid-price movement from different prediction horizons.\\
3. The DeepLOB-Attention and DeepLOB-Seq2Seq performance is comparable to DeepLOB but better than DeepLOB, especially in the long prediction horizon. This result accords with the result in the previous paper \cite{LOBs2s}, which proves the correctness of the result.\\
4. The Autoformer gets the second-best result in prediction horizon 30. Although it is not the best model, it still means that Autoformer is usable for the time series prediction, and its time decomposition structure is adequate. The shortcoming of Autoformer and the latest FEDformer is that they are huge models compared to LSTM, and they need to be fine-tuned to work well in a specific task. In contrast, LSTM-based models’ sizes are much smaller and do not need much hyper-parameters tuning. More analysis of the efficiency will be discussed in the Section~\ref{efficiency}.\\\par
To summarize, combing the results of this task and previous tasks, LSTM-based models generally have their advantage in financial time series for their robustness and good compatibility. Although the Transformer-based model is large and complicated to tune and requires a long training time, they are still usable for the forecasting task. Furthermore, the research of Transformer-based method in time series prediction is meaningful because the time decomposition method from Autoformer contributes back to the original LSTM model.

\begin{figure}[hbt!]
    \centering
    \includegraphics[scale=0.4]{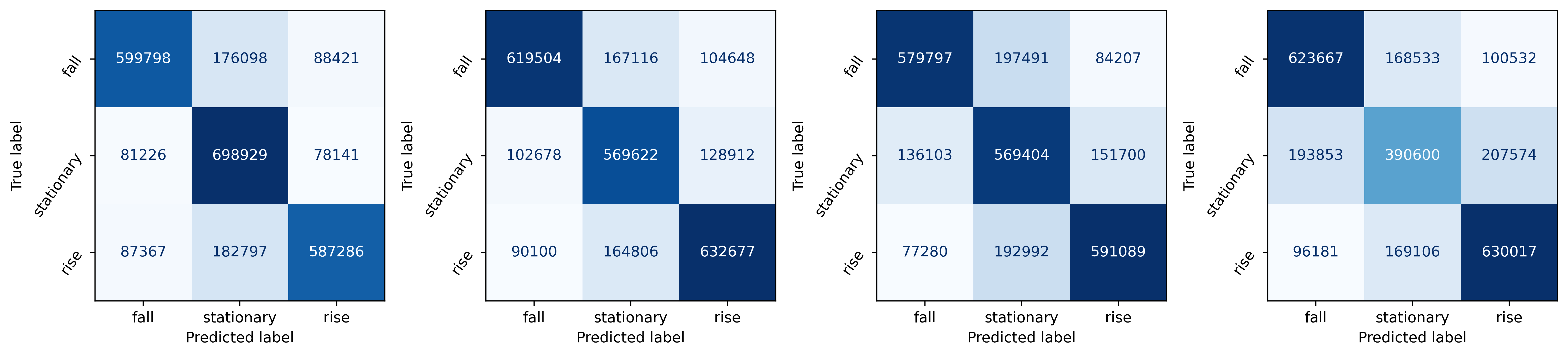}
    \caption{Confusion matrix of DLSTM model with prediction horizon 20, 30, 50, 100.}
    \label{fig.confusion}
\end{figure} 

\subsubsection{Simple Trading Simulation without transaction cost}

In order to show the models and their predictions are practical and useful in trading, a simple trading simulation (backtesting) is designed. Three models with good classification metrics performance are chosen for comparison: DLSTM, DeepLOB \cite{DeepLOB}, Autoformer \cite{Autoformer}. Canonical LSTM \cite{lstm} and Vanilla Transformer \cite{attention}  are used as baselines. The three-day test set is used for this trading simulation. To make a fair comparison among models, the trading simulation follows the simple setup in the previous work \cite{DeepLOB}:
The number of shares pre-trade $\mu$ (volume) is set to one. At each timestamp, the model will predict the price movement ($0$: fall, $1$: stationary, $2$: rise) as a trading signal. When the prediction is $2$, enter the long position, and the position is held until it encounters $0$. The same rule is applied to the short position when the prediction is $0$, and only one direction of position can exist in this simulation trading. A delay is set between the prediction and the order execution to simulate the high-frequency trading latency. For example, assume the model generates a prediction $2$ at time $t$, $\mu$ shares will be bought at time $t+5$.
\begin{table}[hbt!]
\centering
\resizebox{\textwidth}{!}{%
\begin{tabular}{@{}c|cc|cc|cc|cc@{}}
\toprule
Forecast Horizon &
  \multicolumn{2}{c|}{Prediction Horizon = 20} &
  \multicolumn{2}{c|}{Prediction Horizon = 30} &
  \multicolumn{2}{c|}{Prediction Horizon =50} &
  \multicolumn{2}{c}{Prediction Horizon=100} \\ \midrule
Model &
  CPR &
  SR &
  CPR &
  SR &
  CPR &
  SR &
  CPR &
  SR \\ \midrule
LSTM &
  {\color[HTML]{B22222} \textbf{15.396}} &
  {\color[HTML]{333333} 51.489} &
  {\color[HTML]{000000} 12.458} &
  {\color[HTML]{006494} {\ul 41.411}} &
  {\color[HTML]{B22222} \textbf{8.484}} &
  {\color[HTML]{B22222} \textbf{28.817}} &
  {\color[HTML]{B22222} \textbf{4.914}} &
  {\color[HTML]{B22222} \textbf{20.941}} \\
DLSTM &
  {\color[HTML]{006494} {\ul 14.966}} &
  {\color[HTML]{333333} 46.949} &
  {\color[HTML]{333333} 12.634} &
  37.432 &
  {\color[HTML]{333333} 6.194} &
  {\color[HTML]{333333} 22.027} &
  {\color[HTML]{000000} 3.215} &
  {\color[HTML]{006494} {\ul 16.346}} \\
DeepLOB &
  13.859 &
  {\color[HTML]{006494} {\ul 56.094}} &
  {\color[HTML]{B22222} \textbf{12.789}} &
  {\color[HTML]{B22222} \textbf{42.567}} &
  {\color[HTML]{333333} 5.726} &
  {\color[HTML]{333333} 21.014} &
  2.646 &
  {\color[HTML]{000000} 14.992} \\
Transformer &
  14.553 &
  {\color[HTML]{B22222} \textbf{59.995}} &
  {\color[HTML]{006494} {\ul 12.737}} &
  {\color[HTML]{000000} 41.044} &
  {\color[HTML]{000000} 6.896} &
  {\color[HTML]{006494} {\ul 28.147}} &
  2.859 &
  {\color[HTML]{000000} 16.981} \\
Autoformer &
  9.942 &
  32.688 &
  {\color[HTML]{000000} 8.617} &
  30.576 &
  {\color[HTML]{006494} {\ul 8.214}} &
  {\color[HTML]{000000} 25.882} &
  {\color[HTML]{006494} {\ul 3.620}} &
  17.765 \\ \bottomrule
\end{tabular}%
}
\caption{Cumulative price returns and annualized sharpe ratio of different models.}
\label{table:3}
\end{table}\\
Several assumptions are made for the simulation trading:\\
1) Since the trading product is in cryptocurrency exchange, the trading volume is considerable sufficient in the market, which means the simulated trades will not have a market impact.\\
2) The focus of this part of the experiment is to show the practicality of the prediction result and make relative comparisons among models instead of inventing a fully developed high-frequency trading strategy. Industrial HFT trading strategies usually require the combination of different prediction signals and precise entry exit rules \cite{DeepLOB}. For simplicity, the order is assumed to be executed at the mid-price without transaction cost. 
\begin{figure}[h]
    \centering
    \includegraphics[scale=0.1]{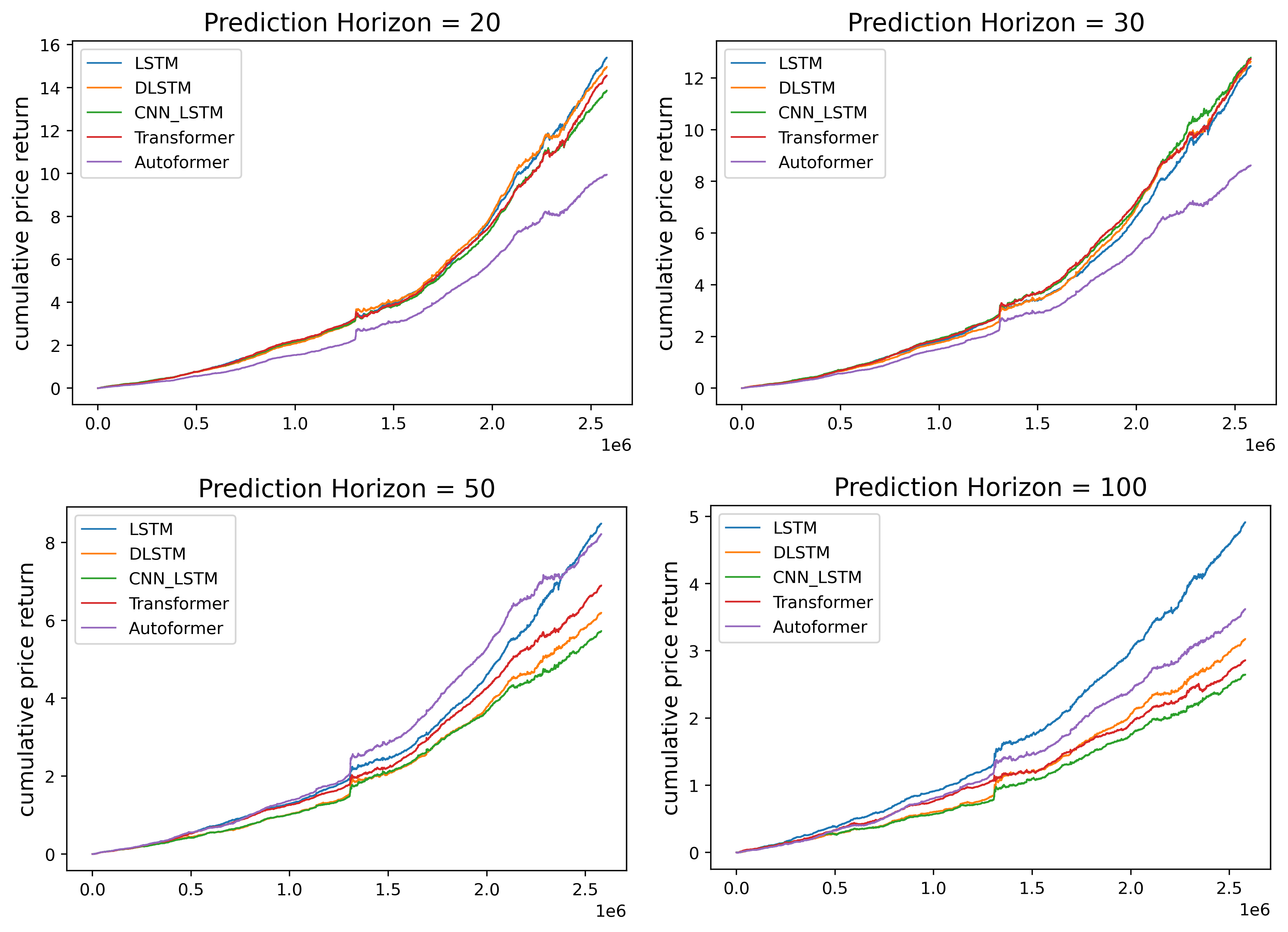}
    \caption{Cumulative return curve for different models in trading simulation.}
    \label{fig.backtest}
\end{figure} \\
As displayed in Table \ref{table:3} and Figure \ref{fig.backtest}, each model’s profitability is presented. The performance of simulated trading is evaluated by cumulative price return (CPR) and the Annualized Sharpe Ratio (SR). The CPR is formulated by:
\begin{equation}
    C P R=\sum_{1}^{t} s * \mu *\left(p_{\text {mid }}^{\text {holding }, t}-p_{\text {mid }}^{\text {settlement }, t}\right)
\end{equation}
where $s$ is the trading position, which is $1$ for long position and $-1$ for short position. $\mu$ is the number of shares.\\ And the Sharpe Ratio is calculated by:
\begin{equation}
    S R=\sqrt{365} \times \frac{\text { Average }(\text { daily } C P R)}{\text { StandardDeviation }(\text { daily } C P R)}
\end{equation}
The value of annualized SR is enormous because the assumptions mentioned above are not realistic for practical trading.\\
Based on the results, LSTM based-model’s performance in simulated trading is generally better than Transformer-based model. The canonical LSTM model achieves highest CPR and SR in prediction horizon 20 and 30 and DeepLOB has the best performance in prediction horizon 50. For DLSTM, it has comparable performance to canonical LSTM and DeepLOB model. This result shows that the prediction result from LSTM-based models are robust and practical for trading. Autoformer’s CPR is the lowest in prediction horizon 20 and 30. The state-of-the-art Autoformer sometimes has even worse performance than the vanilla Transformer in simulated trading, although it obtains a better classification metrics. To summarize, LSTM-based models are relatively the better models for electronic trading.

\subsubsection{Simple Trading Simulation with transaction cost}

\begin{figure}[hbt!]
    \centering
    \includegraphics[scale=0.45]{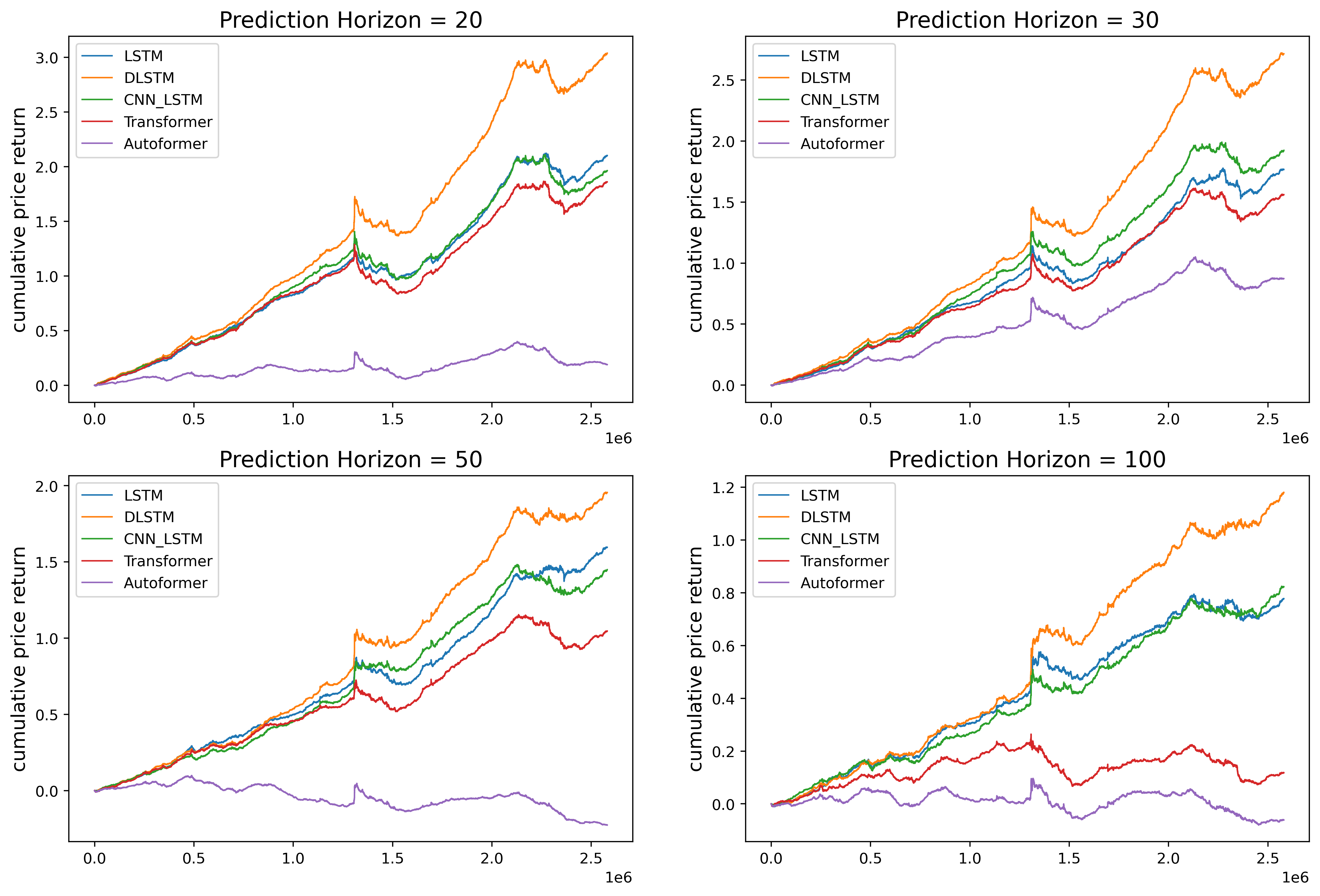}
    \caption{Cumulative return curve under $0.002\%$ transaction cost.}
    \label{fig.backtest_fee}
\end{figure}

In the real-world market, all operations, including buying or selling, need a commission fee, and sometimes the transaction cost might outweight the return. This section will introduce a hypothetical transaction cost of $0.002\%$ to further compare the robustness among different models. The results are shown in Table \ref{table.backtest_tx} and Figure \ref{fig.backtest_fee}.
\begin{table}[hbt!]
\centering
\resizebox{\textwidth}{!}{%
\begin{tabular}{@{}c|cc|cc|cc|cc@{}}
\toprule
Forecast Horizon &
  \multicolumn{2}{c|}{Prediction Horizon = 20} &
  \multicolumn{2}{c|}{Prediction Horizon = 30} &
  \multicolumn{2}{c|}{Prediction Horizon =50} &
  \multicolumn{2}{c}{Prediction Horizon=100} \\ \midrule
Model &
  CPR &
  SR &
  CPR &
  SR &
  CPR &
  SR &
  CPR &
  SR \\ \midrule
LSTM &
  {\color[HTML]{006494} {\ul 2.102}} &
  {\color[HTML]{006494} {\ul 15.160}} &
  1.767 &
  12.429 &
  {\color[HTML]{006494} {\ul 1.596}} &
  {\color[HTML]{006494} {\ul 11.536}} &
  0.778 &
  6.014 \\
DLSTM &
  {\color[HTML]{B22222} \textbf{3.039}} &
  {\color[HTML]{B22222} \textbf{19.962}} &
  {\color[HTML]{B22222} \textbf{2.716}} &
  {\color[HTML]{B22222} \textbf{16.523}} &
  {\color[HTML]{B22222} \textbf{1.957}} &
  {\color[HTML]{B22222} \textbf{12.359}} &
  {\color[HTML]{B22222} \textbf{1.180}} &
  {\color[HTML]{B22222} \textbf{9.811}} \\
DeepLOB &
  1.964 &
  15.082 &
  {\color[HTML]{006494} {\ul 1.924}} &
  {\color[HTML]{006494} {\ul 13.128}} &
  1.450 &
  10.273 &
  {\color[HTML]{006494} {\ul 0.823}} &
  {\color[HTML]{006494} {\ul 7.993}} \\
Transformer &
  1.860 &
  13.894 &
  1.561 &
  10.917 &
  1.047 &
  6.612 &
  0.118 &
  -23.496 \\
Autoformer &
  0.189 &
  -8.704 &
  0.873 &
  5.118 &
  -0.225 &
  -9.193 &
  -0.061 &
  -14.835 \\ \bottomrule
\end{tabular}%
}
\caption{Cumulative price returns and annualized sharpe ratio of different models under $0.002\%$ transaction cost.}
\label{table.backtest_tx}
\end{table}

From the table, DLSTM has the highest CPRs and SRs for all the prediction horizons, outperforming all other models. This shows DLSTM's strong profitability and robustness against the risk brought by the transaction cost. LSTM-based methods' performance is generally better than Transformer-based methods. Canonical LSTM and DeepLOB achieve the second-best CPRs and SRs in different prediction horizons. This indicates that the LSTM-based model's prediction results are more practical and effective in electronic trading. Interestingly, Transformer-based models' performance drops significantly under the transaction cost. The state-of-the-art Autoformer produces even less profit than vanilla Transformer, yielding negative CPRs and SRs in prediction horizon 50 and 100, although its prediction classification metrics is better than Transformer.\\

\begin{figure}[hbt!]
    \centering
    \includegraphics[scale=0.42]{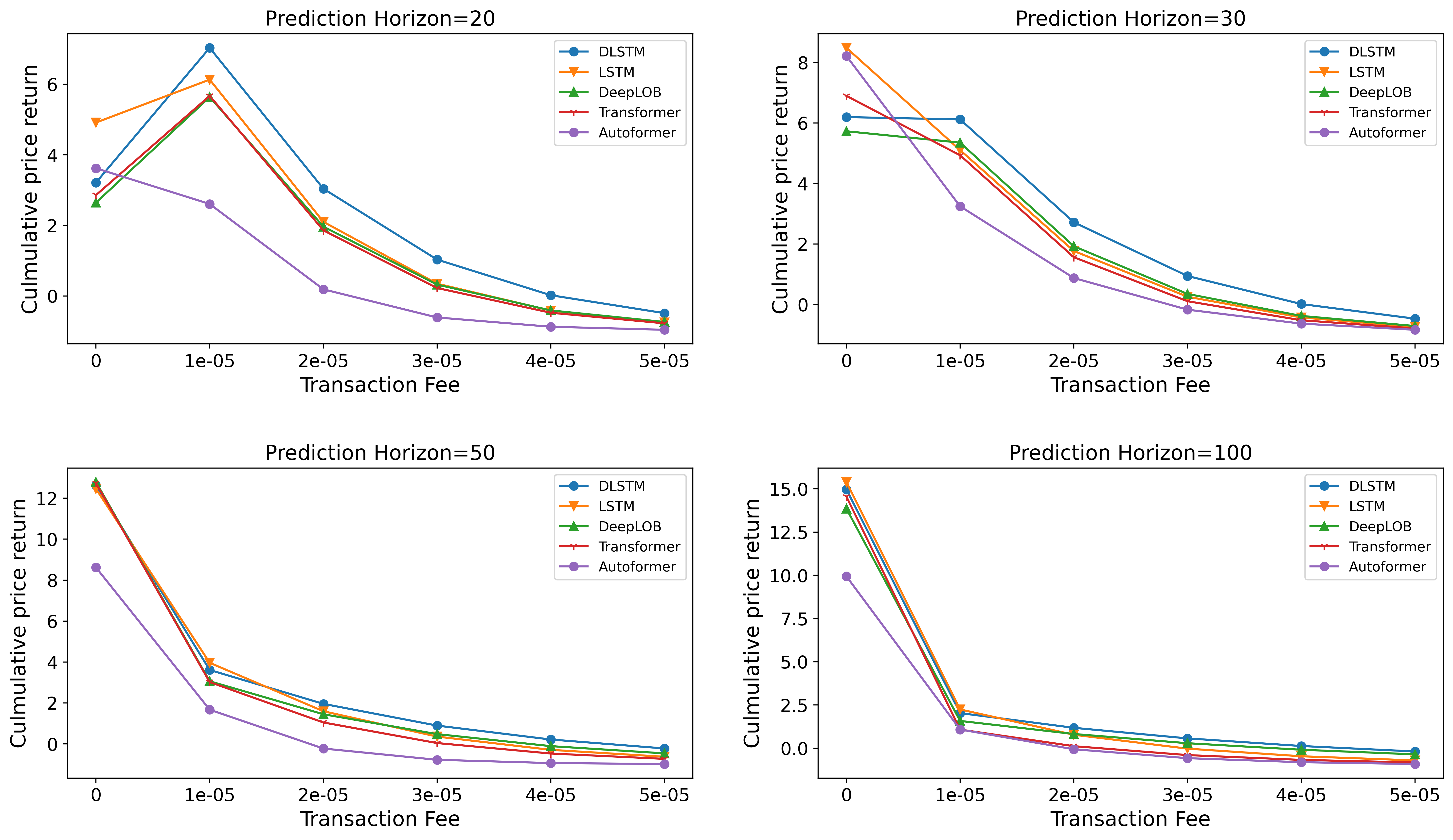}
    \caption{Cumulative price return change with increasing transaction cost (back testing period: $2022.07.12$ (inclusive) to $2022.07.14$ (inclusive)).}
    \label{fig.cpr_tx}
\end{figure}

\begin{figure}[hbt!]
    \centering
    \includegraphics[scale=0.42]{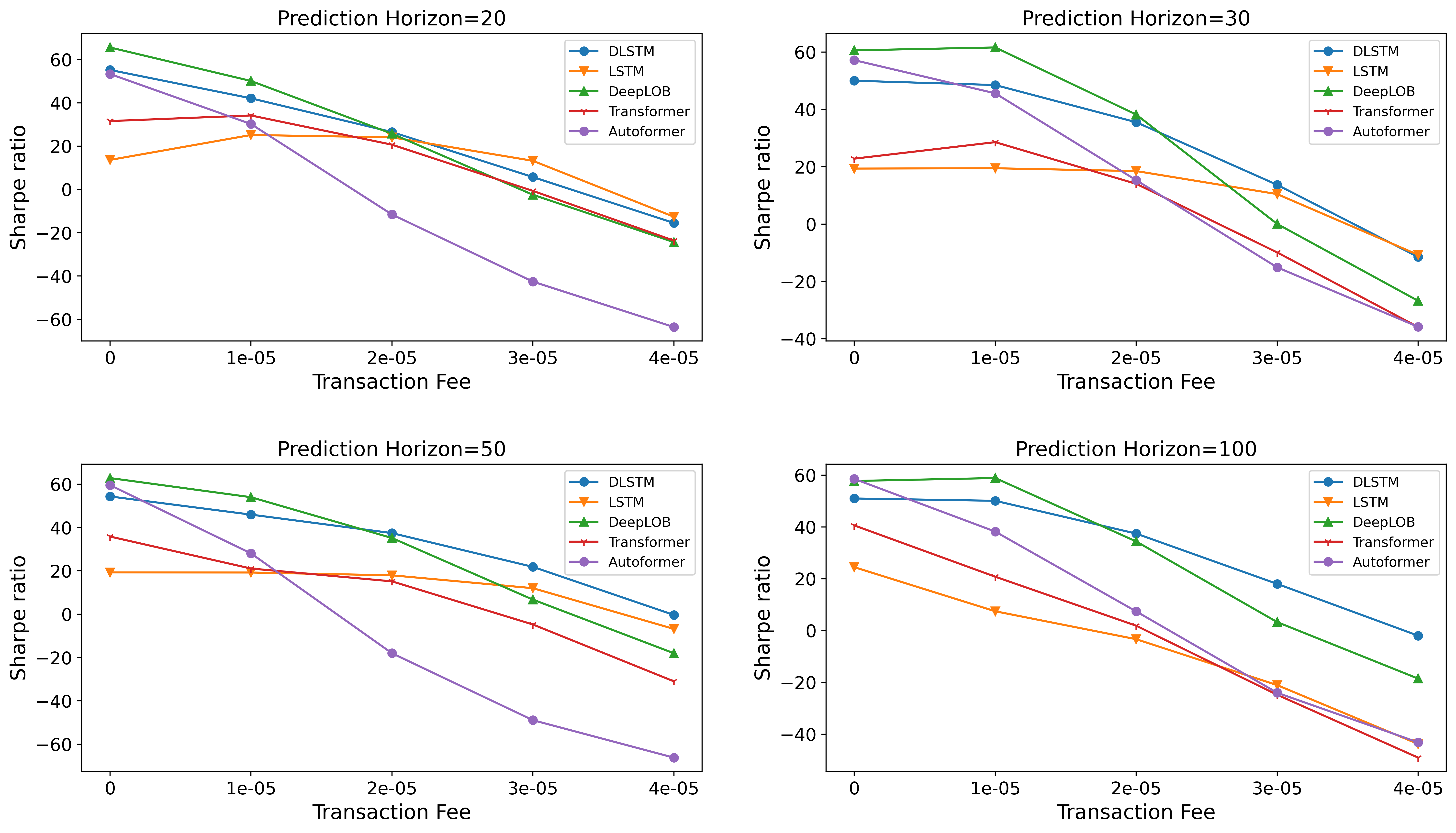}
    \caption{Sharpe ratio change with increasing transaction cost (back testings period: $2022.07.13$ (inclusive) to $2022.07.24$ (inclusive)).}
    \label{fig.sr_tx}
\end{figure}

To further investigate the impact of transaction cost on simulated trading. An experiment is extended to see how the CPR and SR change as the transaction cost increase. The experiment on CPR is done on the three-day test set from $2022.07.12$ (inclusive) to $2022.07.14$ (inclusive) as mentioned in Section~\ref{task3_setting}. The experiment on SR has a longer backtesting period ranging from $2022.07.13$ (inclusive) to $2022.07.24$ (inclusive) because annualized Sharpe Ratio is calculated based on daily CPR, so using a longer backtesting period can produce a more accurate SR. The experiment results are shown in Figure \ref{fig.cpr_tx} and Figure \ref{fig.sr_tx}. 
\\In terms of the CPR, all models’ CPR decreases as the transaction cost increases. For the prediction horizon $20$ and $30$, DLSTM still outperforms other models when the transaction cost increases. All the models generate comparable CPR except the Autoformer in the prediction horizon $50$ and $100$. Autoformer’s CPR is the lowest in all prediction horizons for most transaction cost settings. Regarding the SR, all models’ SR decrease as the transaction cost increases. However, DLSTM maintains a higher SR than other models as the transaction cost increases. At the same time, for the Autoformer, its SR drops significantly and even becomes the lowest as the transaction cost increases. Overall, DLSTM keeps its profitability and robustness for different transaction costs, while the Transformer-based method’s performance can be largely affected by the transaction cost. This result further indicates that the LSTM-based method is superior for electronic trading.

\subsubsection{Efficiency Comparison on Transformers versus LSTM} \label{efficiency}

\begin{table}[hbt!]
\centering
\resizebox{\textwidth}{!}{%
\begin{tabular}{@{}l|cccc|ccc@{}}
\toprule
Method &
  \multicolumn{1}{l}{MACs} &
  \multicolumn{1}{l}{Parameter} &
  \multicolumn{1}{l}{Time} &
  \multicolumn{1}{l|}{Memory} &
  \multicolumn{1}{l}{Time} &
  \multicolumn{1}{l}{Memory} &
  \multicolumn{1}{l}{Test Step} \\ \midrule
DLSTM       & 6.72 M  & 193.9 k  & 4.4ms  & 1404MiB & $O(L)$     & $O(L)$     & 1 \\ \midrule
DeepLOB     & 36.42 M & 143.91 k & 6.3ms  & 2250MiB & $O(L)$     & $O(L)$     & 1 \\
Transformer & 1.25 G  & 10.64 M  & 17.7ms & 3534MiB & $O(L^2)$   & $O(L^2)$   & 1 \\
Reformer    & 1.17 G  & 5.84 M   & 23.6ms & 4966MiB & $O(LlogL)$ & $O(L^2)$   & 1 \\
Informer    & 1.15 G  & 11.43 M  & 22.1ms & 4361MiB & $O(LlogL)$ & $O(LlogL)$ & 1 \\
Autoformer  & 1.25 G  & 10.64M   & 75.2ms & 5394MiB & $O(L)$     & $O(L)$     & 1 \\
FEDformer   & 1.25 G  & 16.47 M  & 38.3ms & 3556MiB & $O(L)$     & $O(L)$     & 1 \\ \bottomrule
\end{tabular}%
}
\caption{Efficiency comparison of Transformer-based and LSTM-based models on price movement prediction. MACs represent the number of Multiply-accumulate operations.}
\label{table:eff}
\end{table}
The efficiency comparison of Transformer-based and LSTM-based models on price movement prediction is shown in Table \ref{table:eff}. The comparison is separated into two parts, the left panel is the practical efficiency, and the right is the theoretical efficiency. 
The latest transformer-based models have a focus on lowering the time and memory complexity. Autoformer and FEDformer claim to achieve $O(L)$ time and memory complexity in theory. However, their actual inference time and memory consumption are higher than the vanilla transformer models because of their complex design. For the training process, it usually takes more than $12$ hours to train an Autoformer and FEDformer model, even with the cutting-edge GPU device (e.g., 24GB NVIDIA RTX 3090 GPU is used here), which is not efficient to retrain the model on new data. The researchers should reconsider the focus of the Transformer-based model on time series application. The time and memory complexity is not a big threshold for the vanilla Transformer, where its inference speed and memory consumption is acceptable depending on today’s computing power. \\\par

The LSTM-based model has higher efficiency than the Transformer-based model for its low inference time and small model size, where its theoretical efficiency corresponds to its practical efficiency. This gives the LSTM-based model advantage in high-frequency trading, which requires fast execution speed. This also again emphasizes that LSTM-based models are the better model in electronic trading.

\section{Conclusion and Future work} 
This study systematically compares LSTM-based and Transformer-based models among three financial time series prediction tasks based on cryptocurrency LOB data. The first task is to predict the LOB mid-price. FEDformer and Autoformer have less error than other models, and LSTM is still a strong model that surpasses Informer, Reformer and vanilla Transformer. Although the mid-price prediction error is low, the quality of the mid-price prediction result is far from sufficient for practical use in high-frequency trading. The second task is to predict LOB mid-price difference. LSTM-based methods show their robustness in time series prediction and perform better than Transformer-based models, which reach the highest $11.5\%$ $R^{2}$ in around $10$ prediction steps. State-of-the-art Autoformer and FEDformer are limited in this task because their time decomposition architecture can not handle the difference sequence. However, in a separate study~\cite{Barez2023}, it was shown that custom transformer configurations can outperform the standard transformers.\\\par

The last task is to predict the LOB mid-price movement. New architecture for the Transformer-based model is designed for adapting the classification task. A new DLSTM model is proposed combining the merits of LSTM and time decomposition architecture from Autoformer. DLSTM outperforms all other models in classification metrics, and Autoformer shows comparable performance to LSTM-based models. A simple trading simulation is done to verify the practicality of the prediction. LSTM-based models have overall better performance than Transformer-based models and DLSTM model beats all other models under the transaction cost.\\\par

In conclusion, based on all the experiments on three different tasks, the Transformer-based model can only outperform LSTM-based models by a large margin in terms of the limited metrics for mid-price prediction. In comparison, the LSTM-based model is still dominant in the later two tasks, so LSTM-based models are generally the better model in financial time series prediction for electronic trading.\\\par

For future research, applying LSTM-based and Transformer-based models in Deep Reinforcement Learning (DRL) can be a proper direction. A complete high-frequency trading strategy usually requires the combination of different prediction signals and needs an experienced trader to control the take-profit and stop-loss. In this case, using DRL to generate the optimal trading strategy directly can get us one step closer to the actual trading. 


\clearpage
\renewcommand\bibname{References}

\bibliographystyle{unsrtnat}

\begin{thebibliography}{63}
\providecommand{\natexlab}[1]{#1}
\providecommand{\url}[1]{\texttt{#1}}
\expandafter\ifx\csname urlstyle\endcsname\relax
  \providecommand{\doi}[1]{doi: #1}\else
  \providecommand{\doi}{doi: \begingroup \urlstyle{rm}\Url}\fi

\bibitem[Fama(1970)]{EMH}
Eugene~F. Fama.
\newblock Efficient capital markets: A review of theory and empirical work.
\newblock \emph{The Journal of finance (New York)}, 25\penalty0 (2):\penalty0
  383--, 1970.
\newblock ISSN 0022-1082.

\bibitem[Murphy(1999)]{TA}
John~J. Murphy.
\newblock \emph{Study guide for Technical analysis of the financial markets : a
  comprehensive guide to trading methods and applications}.
\newblock New York Institute of Finance, New York, 1999.
\newblock ISBN 0735200653.

\bibitem[Brown(2012)]{alma991000618293301591}
Constance~M. Brown.
\newblock \emph{Mastering elliott wave principle elementary concepts, wave
  patterns, and practice exercises}.
\newblock Bloomberg financial series. Wiley, Hoboken, N.J, 1st edition edition,
  2012.
\newblock ISBN 1-280-67304-4.

\bibitem[Boroden(2008)]{boroden2008fibonacci}
Carolyn Boroden.
\newblock \emph{Fibonacci trading: how to master the time and price advantage}.
\newblock Mcgraw-hill New York, NY, 2008.

\bibitem[Ruppert(2015)]{alma991000193490801591}
David. Ruppert.
\newblock \emph{Statistics and Data Analysis for Financial Engineering with R
  examples}.
\newblock Springer Texts in Statistics. Springer New York, New York, NY, 2nd
  ed. 2015. edition, 2015.
\newblock ISBN 1-4939-2614-4.

\bibitem[Ariyo et~al.(2014)Ariyo, Adewumi, and Ayo]{7046047}
Adebiyi~A. Ariyo, Adewumi~O. Adewumi, and Charles~K. Ayo.
\newblock Stock price prediction using the arima model.
\newblock In \emph{2014 UKSim-AMSS 16th International Conference on Computer
  Modelling and Simulation}, pages 106--112, 2014.
\newblock \doi{10.1109/UKSim.2014.67}.

\bibitem[Carbune et~al.(2019)Carbune, Gonnet, Deselaers, Rowley, Daryin, Calvo,
  Wang, Keysers, Feuz, and Gervais]{DBLP:journals/corr/abs-1902-10525}
Victor Carbune, Pedro Gonnet, Thomas Deselaers, Henry~A. Rowley, Alexander~N.
  Daryin, Marcos Calvo, Li{-}Lun Wang, Daniel Keysers, Sandro Feuz, and
  Philippe Gervais.
\newblock Fast multi-language lstm-based online handwriting recognition.
\newblock \emph{CoRR}, abs/1902.10525, 2019.
\newblock URL \url{http://arxiv.org/abs/1902.10525}.

\bibitem[Soltau et~al.(2016)Soltau, Liao, and
  Sak]{https://doi.org/10.48550/arxiv.1610.09975}
Hagen Soltau, Hank Liao, and Hasim Sak.
\newblock Neural speech recognizer: Acoustic-to-word lstm model for large
  vocabulary speech recognition, 2016.
\newblock URL \url{https://arxiv.org/abs/1610.09975}.

\bibitem[Vaswani et~al.(2017)Vaswani, Shazeer, Parmar, Uszkoreit, Jones, Gomez,
  Kaiser, and Polosukhin]{attention}
Ashish Vaswani, Noam Shazeer, Niki Parmar, Jakob Uszkoreit, Llion Jones,
  Aidan~N. Gomez, Lukasz Kaiser, and Illia Polosukhin.
\newblock Attention is all you need.
\newblock \emph{CoRR}, abs/1706.03762, 2017.
\newblock URL \url{http://arxiv.org/abs/1706.03762}.

\bibitem[Sirignano and Cont(2018)]{LSTMLOB}
Justin Sirignano and Rama Cont.
\newblock Universal features of price formation in financial markets:
  perspectives from deep learning, 2018.
\newblock URL \url{https://arxiv.org/abs/1803.06917}.

\bibitem[Zhang et~al.(2019{\natexlab{a}})Zhang, Zohren, and Roberts]{DeepLOB}
Zihao Zhang, Stefan Zohren, and Stephen Roberts.
\newblock {DeepLOB}: Deep convolutional neural networks for limit order books.
\newblock \emph{{IEEE} Transactions on Signal Processing}, 67\penalty0
  (11):\penalty0 3001--3012, jun 2019{\natexlab{a}}.
\newblock \doi{10.1109/tsp.2019.2907260}.
\newblock URL \url{https://doi.org/10.1109%2Ftsp.2019.2907260}.

\bibitem[Zhang and Zohren(2021)]{LOBs2s}
Zihao Zhang and Stefan Zohren.
\newblock Multi-horizon forecasting for limit order books: Novel deep learning
  approaches and hardware acceleration using intelligent processing units.
\newblock \emph{CoRR}, abs/2105.10430, 2021.
\newblock URL \url{https://arxiv.org/abs/2105.10430}.

\bibitem[Tsantekidis et~al.(2018)Tsantekidis, Passalis, Tefas, Kanniainen,
  Gabbouj, and Iosifidis]{https://doi.org/10.48550/arxiv.1810.09965}
Avraam Tsantekidis, Nikolaos Passalis, Anastasios Tefas, Juho Kanniainen,
  Moncef Gabbouj, and Alexandros Iosifidis.
\newblock Using deep learning for price prediction by exploiting stationary
  limit order book features, 2018.
\newblock URL \url{https://arxiv.org/abs/1810.09965}.

\bibitem[Kolm et~al.(2021)Kolm, Turiel, and Westray]{Kolm2021DeepOF}
Petter~N. Kolm, Jeremy~D. Turiel, and Nicholas Westray.
\newblock Deep order flow imbalance: Extracting alpha at multiple horizons from
  the limit order book.
\newblock \emph{Econometric Modeling: Capital Markets - Portfolio Theory
  eJournal}, 2021.

\bibitem[Roondiwala et~al.(2017)Roondiwala, Patel, and Varma]{prdictLSTM}
Murtaza Roondiwala, Harshal Patel, and Shraddha Varma.
\newblock Predicting stock prices using lstm.
\newblock \emph{International Journal of Science and Research (IJSR)}, 6, 04
  2017.
\newblock \doi{10.21275/ART20172755}.

\bibitem[Cao et~al.(2019)Cao, Li, and Li]{CAO2019127}
Jian Cao, Zhi Li, and Jian Li.
\newblock Financial time series forecasting model based on ceemdan and lstm.
\newblock \emph{Physica A: Statistical Mechanics and its Applications},
  519:\penalty0 127--139, 2019.
\newblock ISSN 0378-4371.
\newblock \doi{https://doi.org/10.1016/j.physa.2018.11.061}.
\newblock URL
  \url{https://www.sciencedirect.com/science/article/pii/S0378437118314985}.

\bibitem[Bao et~al.(2017)Bao, Yue, and Rao]{10.1371/journal.pone.0180944}
Wei Bao, Jun Yue, and Yulei Rao.
\newblock A deep learning framework for financial time series using stacked
  autoencoders and long-short term memory.
\newblock \emph{PLOS ONE}, 12\penalty0 (7):\penalty0 1--24, 07 2017.
\newblock \doi{10.1371/journal.pone.0180944}.
\newblock URL \url{https://doi.org/10.1371/journal.pone.0180944}.

\bibitem[Selvin et~al.(2017)Selvin, Vinayakumar, Gopalakrishnan, Menon, and
  Soman]{8126078}
Sreelekshmy Selvin, R~Vinayakumar, E.~A Gopalakrishnan, Vijay~Krishna Menon,
  and K.~P. Soman.
\newblock Stock price prediction using lstm, rnn and cnn-sliding window model.
\newblock In \emph{2017 International Conference on Advances in Computing,
  Communications and Informatics (ICACCI)}, pages 1643--1647, 2017.
\newblock \doi{10.1109/ICACCI.2017.8126078}.

\bibitem[Fischer and Krauss(2018)]{FISCHER2018654}
Thomas Fischer and Christopher Krauss.
\newblock Deep learning with long short-term memory networks for financial
  market predictions.
\newblock \emph{European Journal of Operational Research}, 270\penalty0
  (2):\penalty0 654--669, 2018.
\newblock ISSN 0377-2217.
\newblock \doi{https://doi.org/10.1016/j.ejor.2017.11.054}.
\newblock URL
  \url{https://www.sciencedirect.com/science/article/pii/S0377221717310652}.

\bibitem[Siami{-}Namini et~al.(2019)Siami{-}Namini, Tavakoli, and
  Namin]{DBLP:journals/corr/abs-1911-09512}
Sima Siami{-}Namini, Neda Tavakoli, and Akbar~Siami Namin.
\newblock A comparative analysis of forecasting financial time series using
  arima, lstm, and bilstm.
\newblock \emph{CoRR}, abs/1911.09512, 2019.
\newblock URL \url{http://arxiv.org/abs/1911.09512}.

\bibitem[Kim and Kang(2019)]{AttenLSTM}
Sangyeon Kim and Myungjoo Kang.
\newblock Financial series prediction using attention lstm, 2019.
\newblock URL \url{https://arxiv.org/abs/1902.10877}.

\bibitem[Zhang et~al.(2019{\natexlab{b}})Zhang, Liang, Zhiyuli, Zhang, Xu, and
  Wu]{Zhang_2019}
Xuan Zhang, Xun Liang, Aakas Zhiyuli, Shusen Zhang, Rui Xu, and Bo~Wu.
\newblock {AT}-{LSTM}: An attention-based {LSTM} model for financial time
  series prediction.
\newblock \emph{{IOP} Conference Series: Materials Science and Engineering},
  569\penalty0 (5):\penalty0 052037, jul 2019{\natexlab{b}}.
\newblock \doi{10.1088/1757-899x/569/5/052037}.
\newblock URL \url{https://doi.org/10.1088/1757-899x/569/5/052037}.

\bibitem[Hu(2021)]{9731073}
Xiaokang Hu.
\newblock Stock price prediction based on temporal fusion transformer.
\newblock In \emph{2021 3rd International Conference on Machine Learning, Big
  Data and Business Intelligence (MLBDBI)}, pages 60--66, 2021.
\newblock \doi{10.1109/MLBDBI54094.2021.00019}.

\bibitem[Sridhar and Sanagavarapu(2021)]{9538640}
Sashank Sridhar and Sowmya Sanagavarapu.
\newblock Multi-head self-attention transformer for dogecoin price prediction.
\newblock In \emph{2021 14th International Conference on Human System
  Interaction (HSI)}, pages 1--6, 2021.
\newblock \doi{10.1109/HSI52170.2021.9538640}.

\bibitem[Sonkiya et~al.(2021)Sonkiya, Bajpai, and Bansal]{predictBERT}
Priyank Sonkiya, Vikas Bajpai, and Anukriti Bansal.
\newblock Stock price prediction using bert and gan, 2021.
\newblock URL \url{https://arxiv.org/abs/2107.09055}.

\bibitem[Lakew et~al.(2018)Lakew, Cettolo, and
  Federico]{https://doi.org/10.48550/arxiv.1806.06957}
Surafel~M. Lakew, Mauro Cettolo, and Marcello Federico.
\newblock A comparison of transformer and recurrent neural networks on
  multilingual neural machine translation, 2018.
\newblock URL \url{https://arxiv.org/abs/1806.06957}.

\bibitem[Karita et~al.(2019)Karita, Chen, Hayashi, Hori, Inaguma, Jiang,
  Someki, Soplin, Yamamoto, Wang, Watanabe, Yoshimura, and Zhang]{Karita_2019}
Shigeki Karita, Nanxin Chen, Tomoki Hayashi, Takaaki Hori, Hirofumi Inaguma,
  Ziyan Jiang, Masao Someki, Nelson Enrique~Yalta Soplin, Ryuichi Yamamoto,
  Xiaofei Wang, Shinji Watanabe, Takenori Yoshimura, and Wangyou Zhang.
\newblock A comparative study on transformer vs {RNN} in speech applications.
\newblock In \emph{2019 {IEEE} Automatic Speech Recognition and Understanding
  Workshop ({ASRU})}. {IEEE}, dec 2019.
\newblock \doi{10.1109/asru46091.2019.9003750}.
\newblock URL \url{https://doi.org/10.1109%2Fasru46091.2019.9003750}.

\bibitem[Wen et~al.(2022)Wen, Zhou, Zhang, Chen, Ma, Yan, and Sun]{SURVEY}
Qingsong Wen, Tian Zhou, Chaoli Zhang, Weiqi Chen, Ziqing Ma, Junchi Yan, and
  Liang Sun.
\newblock Transformers in time series: A survey, 2022.
\newblock URL \url{https://arxiv.org/abs/2202.07125}.

\bibitem[Sutskever et~al.(2014)Sutskever, Vinyals, and Le]{S2S}
Ilya Sutskever, Oriol Vinyals, and Quoc~V. Le.
\newblock Sequence to sequence learning with neural networks, 2014.
\newblock URL \url{https://arxiv.org/abs/1409.3215}.

\bibitem[Brown et~al.(2020)Brown, Mann, Ryder, Subbiah, Kaplan, Dhariwal,
  Neelakantan, Shyam, Sastry, Askell, Agarwal, Herbert-Voss, Krueger, Henighan,
  Child, Ramesh, Ziegler, Wu, Winter, Hesse, Chen, Sigler, Litwin, Gray, Chess,
  Clark, Berner, McCandlish, Radford, Sutskever, and
  Amodei]{https://doi.org/10.48550/arxiv.2005.14165}
Tom~B. Brown, Benjamin Mann, Nick Ryder, Melanie Subbiah, Jared Kaplan,
  Prafulla Dhariwal, Arvind Neelakantan, Pranav Shyam, Girish Sastry, Amanda
  Askell, Sandhini Agarwal, Ariel Herbert-Voss, Gretchen Krueger, Tom Henighan,
  Rewon Child, Aditya Ramesh, Daniel~M. Ziegler, Jeffrey Wu, Clemens Winter,
  Christopher Hesse, Mark Chen, Eric Sigler, Mateusz Litwin, Scott Gray,
  Benjamin Chess, Jack Clark, Christopher Berner, Sam McCandlish, Alec Radford,
  Ilya Sutskever, and Dario Amodei.
\newblock Language models are few-shot learners, 2020.
\newblock URL \url{https://arxiv.org/abs/2005.14165}.

\bibitem[Li et~al.(2019)Li, Jin, Xuan, Zhou, Chen, Wang, and Yan]{LogTrans}
Shiyang Li, Xiaoyong Jin, Yao Xuan, Xiyou Zhou, Wenhu Chen, Yu-Xiang Wang, and
  Xifeng Yan.
\newblock Enhancing the locality and breaking the memory bottleneck of
  transformer on time series forecasting, 2019.
\newblock URL \url{https://arxiv.org/abs/1907.00235}.

\bibitem[Kitaev et~al.(2020)Kitaev, Kaiser, and Levskaya]{Reformer}
Nikita Kitaev, Łukasz Kaiser, and Anselm Levskaya.
\newblock Reformer: The efficient transformer, 2020.
\newblock URL \url{https://arxiv.org/abs/2001.04451}.

\bibitem[Zhou et~al.(2020)Zhou, Zhang, Peng, Zhang, Li, Xiong, and
  Zhang]{Informer}
Haoyi Zhou, Shanghang Zhang, Jieqi Peng, Shuai Zhang, Jianxin Li, Hui Xiong,
  and Wancai Zhang.
\newblock Informer: Beyond efficient transformer for long sequence time-series
  forecasting, 2020.
\newblock URL \url{https://arxiv.org/abs/2012.07436}.

\bibitem[Wu et~al.(2021)Wu, Xu, Wang, and Long]{Autoformer}
Haixu Wu, Jiehui Xu, Jianmin Wang, and Mingsheng Long.
\newblock Autoformer: Decomposition transformers with auto-correlation for
  long-term series forecasting, 2021.
\newblock URL \url{https://arxiv.org/abs/2106.13008}.

\bibitem[Liu et~al.(2022)Liu, Yu, Liao, Li, Lin, Liu, and
  Dustdar]{liu2022pyraformer}
Shizhan Liu, Hang Yu, Cong Liao, Jianguo Li, Weiyao Lin, Alex~X. Liu, and
  Schahram Dustdar.
\newblock Pyraformer: Low-complexity pyramidal attention for long-range time
  series modeling and forecasting.
\newblock In \emph{International Conference on Learning Representations}, 2022.
\newblock URL \url{https://openreview.net/forum?id=0EXmFzUn5I}.

\bibitem[Zhou et~al.(2022{\natexlab{a}})Zhou, Ma, Wen, Wang, Sun, and
  Jin]{FEDFormer}
Tian Zhou, Ziqing Ma, Qingsong Wen, Xue Wang, Liang Sun, and Rong Jin.
\newblock Fedformer: Frequency enhanced decomposed transformer for long-term
  series forecasting, 2022{\natexlab{a}}.
\newblock URL \url{https://arxiv.org/abs/2201.12740}.

\bibitem[Gould et~al.(2010)Gould, Porter, Williams, McDonald, Fenn, and
  Howison]{LOB}
Martin~D. Gould, Mason~A. Porter, Stacy Williams, Mark McDonald, Daniel~J.
  Fenn, and Sam~D. Howison.
\newblock Limit order books, 2010.
\newblock URL \url{https://arxiv.org/abs/1012.0349}.

\bibitem[Tsantekidis et~al.(2017)Tsantekidis, Passalis, Tefas, Kanniainen,
  Gabbouj, and Iosifidis]{8010701}
Avraam Tsantekidis, Nikolaos Passalis, Anastasios Tefas, Juho Kanniainen,
  Moncef Gabbouj, and Alexandros Iosifidis.
\newblock Forecasting stock prices from the limit order book using
  convolutional neural networks.
\newblock In \emph{2017 IEEE 19th Conference on Business Informatics (CBI)},
  volume~01, pages 7--12, 2017.
\newblock \doi{10.1109/CBI.2017.23}.

\bibitem[Hochreiter and Schmidhuber(1997)]{lstm}
Sepp Hochreiter and Jürgen Schmidhuber.
\newblock {Long Short-Term Memory}.
\newblock \emph{Neural Computation}, 9\penalty0 (8):\penalty0 1735--1780, 11
  1997.
\newblock ISSN 0899-7667.
\newblock \doi{10.1162/neco.1997.9.8.1735}.
\newblock URL \url{https://doi.org/10.1162/neco.1997.9.8.1735}.

\bibitem[Goodfellow et~al.(2016)Goodfellow, Bengio, and Courville]{dlbook}
Ian Goodfellow, Yoshua Bengio, and Aaron Courville.
\newblock \emph{Deep Learning}.
\newblock MIT Press, 2016.
\newblock \url{http://www.deeplearningbook.org}.

\bibitem[Rumelhart et~al.(1986)Rumelhart, Hinton, and
  Williams]{Rumelhart1986LearningRB}
David~E. Rumelhart, Geoffrey~E. Hinton, and Ronald~J. Williams.
\newblock Learning representations by back-propagating errors.
\newblock \emph{Nature}, 323:\penalty0 533--536, 1986.

\bibitem[Gers et~al.(1999)Gers, Schmidhuber, and Cummins]{818041}
F.A. Gers, J.~Schmidhuber, and F.~Cummins.
\newblock Learning to forget: continual prediction with lstm.
\newblock In \emph{1999 Ninth International Conference on Artificial Neural
  Networks ICANN 99. (Conf. Publ. No. 470)}, volume~2, pages 850--855 vol.2,
  1999.
\newblock \doi{10.1049/cp:19991218}.

\bibitem[Graves(2013)]{lstmfig1}
Alex Graves.
\newblock Generating sequences with recurrent neural networks, 2013.
\newblock URL \url{https://arxiv.org/abs/1308.0850}.

\bibitem[Oomen and Gatheral(2010)]{micro}
Roel Oomen and Jim Gatheral.
\newblock Zero-intelligence realized variance estimation.
\newblock \emph{Finance and Stochastics}, 14:\penalty0 249--283, 04 2010.
\newblock \doi{10.1007/s00780-009-0120-1}.

\bibitem[Cho et~al.(2014)Cho, van Merrienboer, G{\"{u}}l{\c{c}}ehre, Bougares,
  Schwenk, and Bengio]{seq2seq}
Kyunghyun Cho, Bart van Merrienboer, {\c{C}}aglar G{\"{u}}l{\c{c}}ehre, Fethi
  Bougares, Holger Schwenk, and Yoshua Bengio.
\newblock Learning phrase representations using {RNN} encoder-decoder for
  statistical machine translation.
\newblock \emph{CoRR}, abs/1406.1078, 2014.
\newblock URL \url{http://arxiv.org/abs/1406.1078}.

\bibitem[Luong et~al.(2015)Luong, Pham, and
  Manning]{DBLP:journals/corr/LuongPM15}
Minh{-}Thang Luong, Hieu Pham, and Christopher~D. Manning.
\newblock Effective approaches to attention-based neural machine translation.
\newblock \emph{CoRR}, abs/1508.04025, 2015.
\newblock URL \url{http://arxiv.org/abs/1508.04025}.

\bibitem[Farsani et~al.(2021)Farsani, Pazouki, and Jecei]{TSMTSF}
R~Farsani, Ehsan Pazouki, and Jecei Jecei.
\newblock A transformer self-attention model for time series forecasting.
\newblock \emph{Journal of Electrical and Computer Engineering Innovations},
  9:\penalty0 1--10, 01 2021.
\newblock \doi{10.22061/JECEI.2020.7426.391}.

\bibitem[Zeng et~al.(2022{\natexlab{a}})Zeng, Chen, Zhang, and Xu]{dlinear}
Ailing Zeng, Muxi Chen, Lei Zhang, and Qiang Xu.
\newblock Are transformers effective for time series forecasting?,
  2022{\natexlab{a}}.
\newblock URL \url{https://arxiv.org/abs/2205.13504}.

\bibitem[Taieb and Hyndman(2012)]{IMS}
Souhaib~Ben Taieb and Rob~J Hyndman.
\newblock {Recursive and direct multi-step forecasting: the best of both
  worlds}.
\newblock Monash Econometrics and Business Statistics Working Papers 19/12,
  Monash University, Department of Econometrics and Business Statistics, 2012.
\newblock URL \url{https://ideas.repec.org/p/msh/ebswps/2012-19.html}.

\bibitem[Chevillon(2007)]{DMS}
Guillaume Chevillon.
\newblock Direct multi-step estimation and forecasting.
\newblock \emph{Journal of Economic Surveys}, 21\penalty0 (4):\penalty0
  746--785, 2007.
\newblock \doi{https://doi.org/10.1111/j.1467-6419.2007.00518.x}.
\newblock URL
  \url{https://onlinelibrary.wiley.com/doi/abs/10.1111/j.1467-6419.2007.00518.x}.

\bibitem[Cleveland et~al.(1990)Cleveland, Cleveland, McRae, and
  Terpenning]{STL}
Robert~B. Cleveland, William~S. Cleveland, Jean~E. McRae, and Irma Terpenning.
\newblock Stl: A seasonal-trend decomposition procedure based on loess.
\newblock \emph{Journal of Official Statistics}, 6:\penalty0 3--73, 1990.

\bibitem[Hamilton(1994)]{TS}
James~Douglas Hamilton.
\newblock \emph{Time Series Analysis}.
\newblock Princeton University Press, Princeton, 1994.
\newblock ISBN 0691042896.

\bibitem[Hyndman(2021 - 2021)]{alma991000567577201591}
Rob~J. Hyndman.
\newblock \emph{Forecasting : principles and practice}.
\newblock OTexts, Melbourne, third edition. edition, 2021 - 2021.
\newblock ISBN 9780987507136.

\bibitem[Moscon(2022)]{cryptofeed}
Bryant Moscon.
\newblock cryptofeed.
\newblock \url{https://github.com/bmoscon/cryptofeed}, 2022.

\bibitem[Novotny et~al.(2019)Novotny, Bilokon, Galiotos, and
  D{\'e}l{\`e}ze]{novotny2019machine}
Jan Novotny, Paul~A Bilokon, Aris Galiotos, and Fr{\'e}d{\'e}ric
  D{\'e}l{\`e}ze.
\newblock \emph{Machine Learning and Big Data with kdb+/q}.
\newblock John Wiley \& Sons, 2019.

\bibitem[Wu et~al.(2022)Wu, Xu, Wang, and Long]{Autoformer_repo}
Haixu Wu, Jiehui Xu, Jianmin Wang, and Mingsheng Long.
\newblock Autoformer.
\newblock \url{https://github.com/thuml/Autoformer}, 2022.

\bibitem[Zhou et~al.(2022{\natexlab{b}})Zhou, Ma, Wen, Wang, Sun, and
  Jin]{FEDformer_repo}
Tian Zhou, Ziqing Ma, Qingsong Wen, Xue Wang, Liang Sun, and Rong Jin.
\newblock Fedformer.
\newblock \url{https://github.com/MAZiqing/FEDformer}, 2022{\natexlab{b}}.

\bibitem[Paszke et~al.(2019)Paszke, Gross, Massa, Lerer, Bradbury, Chanan,
  Killeen, Lin, Gimelshein, Antiga, Desmaison, K{\"{o}}pf, Yang, DeVito,
  Raison, Tejani, Chilamkurthy, Steiner, Fang, Bai, and Chintala]{pytorch}
Adam Paszke, Sam Gross, Francisco Massa, Adam Lerer, James Bradbury, Gregory
  Chanan, Trevor Killeen, Zeming Lin, Natalia Gimelshein, Luca Antiga, Alban
  Desmaison, Andreas K{\"{o}}pf, Edward~Z. Yang, Zach DeVito, Martin Raison,
  Alykhan Tejani, Sasank Chilamkurthy, Benoit Steiner, Lu~Fang, Junjie Bai, and
  Soumith Chintala.
\newblock Pytorch: An imperative style, high-performance deep learning library.
\newblock \emph{CoRR}, abs/1912.01703, 2019.
\newblock URL \url{http://arxiv.org/abs/1912.01703}.

\bibitem[Lewis-Beck(1980)]{r_square}
Michael~S. Lewis-Beck.
\newblock \emph{Applied regression : an introduction}.
\newblock Sage university papers series. Quantitative applications in the
  social sciences ; no. 07-022. Sage Publications, Beverly Hills, Calif, 1980.
\newblock ISBN 0803914946.

\bibitem[Zhang(2021{\natexlab{a}})]{DeepLOB_repo}
Zihao Zhang.
\newblock Deeplob-deep-convolutional-neural-networks-for-limit-order-books.
\newblock
  \url{https://github.com/zcakhaa/DeepLOB-Deep-Convolutional-Neural-Networks-for-Limit-Order-Books},
  2021{\natexlab{a}}.

\bibitem[Zhang(2021{\natexlab{b}})]{s2sLOB_repo}
Zihao Zhang.
\newblock Multi-horizon-forecasting-for-limit-order-books.
\newblock
  \url{https://github.com/zcakhaa/Multi-Horizon-Forecasting-for-Limit-Order-Books},
  2021{\natexlab{b}}.

\bibitem[Zeng et~al.(2022{\natexlab{b}})Zeng, Chen, Zhang, and
  Xu]{dlinear_repo}
Ailing Zeng, Muxi Chen, Lei Zhang, and Qiang Xu.
\newblock Ltsf-linear.
\newblock \url{https://github.com/cure-lab/LTSF-Linear}, 2022{\natexlab{b}}.

\bibitem[Barez et~al.(2023)Barez, Bilokon, Gervais, and Lisitsyn]{Barez2023}
Fazl Barez, Paul Bilokon, Arthur Gervais, and Nikita Lisitsyn.
\newblock Exploring the advantages of transformers for high-frequency trading.
\newblock \emph{{SSRN} Electronic Journal}, 2023.
\newblock \doi{10.2139/ssrn.4364833}.

\end{thebibliography}

\appendix

\section{Labelling Details}\label{appendix:detail_label}

\begin{figure}[hbt!]
    \centering
    \includegraphics[scale=0.08]{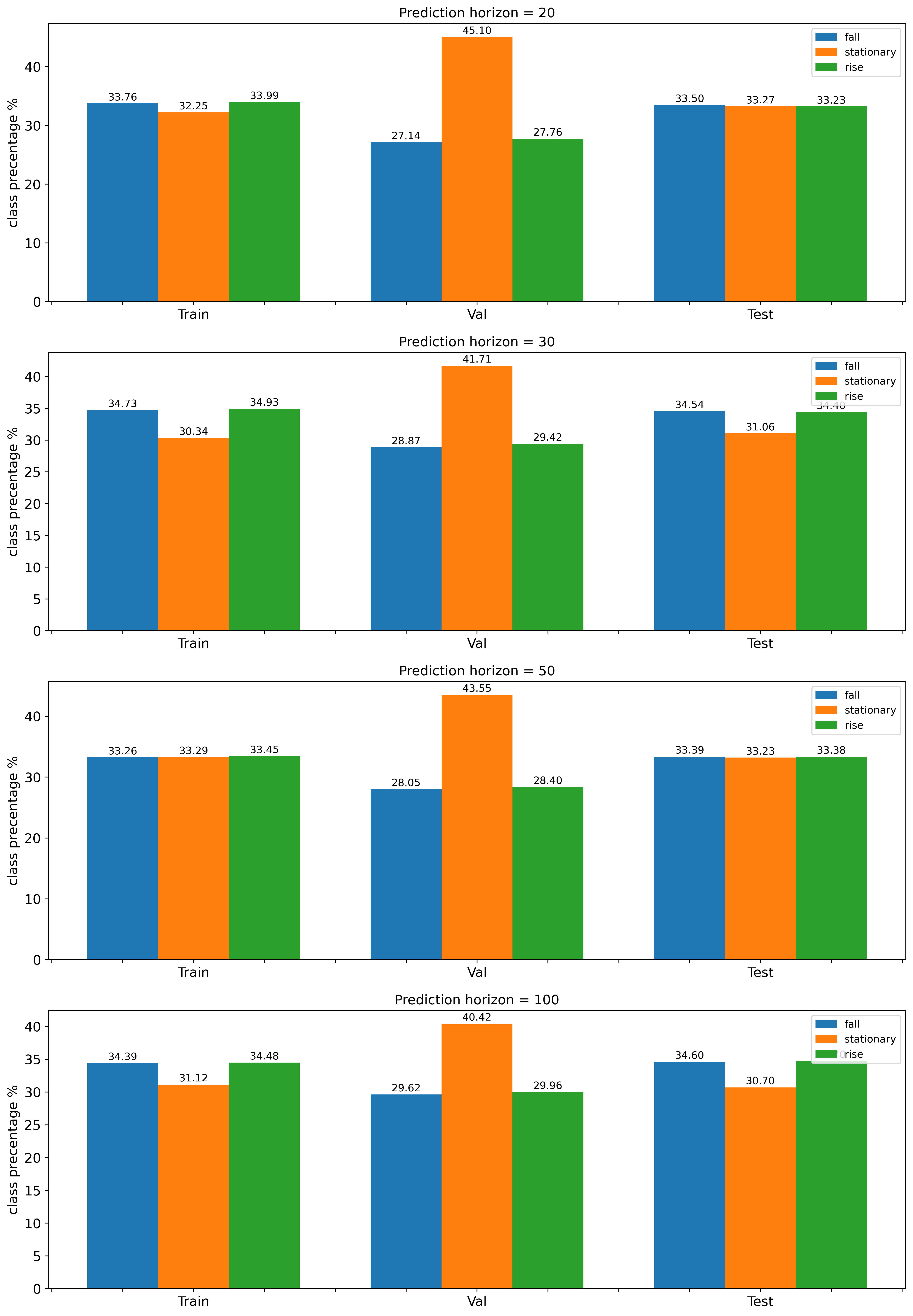}
    \caption{Labelling distribution for different prediction horizon in ETH-USDT dataset}
    \label{fig.all_label}
\end{figure}

As mentioned in Section~\ref{task3}, a threshold $\delta$ needs to be set to decide the corresponding labels. The choice of $\delta$ follows a simple rule, which is to make the labelling roughly balanced. The choice of $\delta$ for different prediction horizon on ETH-USDT dataset is shown in Table \ref{table.all_label} and the distribution of labelling is shown in Figure \ref{fig.all_label}.

\begin{table}[hbt!]
\centering
\begin{tabular}{@{}c|llll@{}}
\toprule
Horizon  & 20   & 30  & 50  & 100  \\ \midrule
$\delta$ & 0.17 & 0.3 & 0.6 & 0.92 \\ \bottomrule
\end{tabular}
\caption{$\delta$ for different prediction horizons for ETH-USDT dataset.(units in $10^{-4}$) }
\label{table.all_label}
\end{table}
\end{document}